\documentclass{amsart}
\usepackage{amsfonts, amssymb, amsmath}

\DeclareSymbolFont{stmry}{U}{stmry}{m}{n}
\DeclareMathDelimiter\llbracket{\mathopen}{stmry}{"4A}{stmry}{"71}    
\DeclareMathDelimiter\rrbracket{\mathclose}{stmry}{"4B}{stmry}{"79}   

\def\d{\partial}

\def\vac{|0\rangle}                            
\def\op{{\mathrm{op}}}

\def\harm{{\mathrm{har}}}

\def\sip{{\mathrm{s.p.}}}                       
\def\CC{\mathbb{C}}       
\def\ZZ{\mathbb{Z}}       
\def\RR{\mathbb{R}}       
\def\NN{\mathbb{N}}       

\def\z{{\mathrm{z}}}      
\def\w{{\mathrm{w}}}
\def\u{{\mathrm{u}}}
\def\p{{\mathrm{p}}}
\def\pp{p}

\def\T{{\mathrm{T}}}

\def\zero{{\mathrm{0}}}

\def\al{\alpha}                         
\def\be{\beta}
\def\ga{\gamma}
\def\Ga{\Gamma}
\def\de{\delta}

\def\io{\iota}

\def\si{\sigma}

\def\ze{\zeta}

\def\sl{{\mathfrak{sl}}}

\DeclareMathOperator{\Res}{Res}

\DeclareMathOperator{\id}{id}

\DeclareMathOperator{\Hom}{Hom}
\DeclareMathOperator{\End}{End}

\def\setcntrs{\setcounter{equation}{0}\setcounter{Theorem}{0}\setcounter{Remark}{0}\setcounter{Definition}{0}\setcounter{Example}{0}\setcounter{Exercise}{0}}

\newcommand{\beq}{\begin{equation}}
\newcommand{\eeq}{\end{equation}}
\newcommand{\beqa}{\begin{eqnarray}}
\newcommand{\eeqa}{\end{eqnarray}}
\newcommand{\nn}{\nonumber \\}

\def\NNN{{\mathbb{Z}_+}}                                           
\def\ZZP{{\widehat{\ZZ}}}                                          
\def\SS{{\mathbb S}}                                             
\def\MM{\overline{M}}                                            
\def\Oo{{\mathrm{O}}}                                            
\def\FF{\phi}                                                    
\def\KK{\mathcal{K}}                                             
\def\di{\partial}                                                
\def\Di{\partial}                                       
\def\spr{\cdot}                                                  
\def\har{\mathfrak{h}}                                           
\def\com{}
\def\hiota{\iota}                                      
\DeclareMathOperator{\RES}{res}                                  
\newcommand{\HR}[1]{H_{#1}}                  
\newcommand{\Txfrac}[2]{\frac{\raisebox{2pt}{$#1$}}{\raisebox{-5pt}{$#2$}}} 

\newcounter{Theorem}
\newcounter{Remark}
\newcounter{Definition}
\newcounter{Example}
\newcounter{Exercise}

\renewcommand{\thesection}{\arabic{section}}

\newcommand{\skp}{
  
  \medskip                                                         
  
  }

\newcommand{\bsec}{}

\newenvironment{Theorem}[1][\bf Theorem \thesection.\arabic{Theorem}]{
        
        \refstepcounter{Theorem}\noindent\textbf{#1.}\it ${}$ ${}$}{}
\newenvironment{Proposition}[1][\bf Proposition \thesection.\arabic{Theorem}]{
        
        \refstepcounter{Theorem}\noindent\textbf{#1.}\it ${}$ ${}$}{}

\newenvironment{Lemma}[1][\bf Lemma \thesection.\arabic{Theorem}]{
        
        \refstepcounter{Theorem}\noindent\textbf{#1.}\it ${}$ ${}$}{}
\newenvironment{Corollary}[1][\bf Corollary \thesection.\arabic{Theorem}]{
        
        \refstepcounter{Theorem}\noindent\textbf{#1.}\it ${}$ ${}$}{}

\newenvironment{Remark}[1][\it Remark \thesection.\arabic{Remark}]{
        
        \refstepcounter{Remark}\noindent\textbf{#1.} }{}
\newenvironment{Example}[1][\bf Example \thesection.\arabic{Example}]{
        
        \refstepcounter{Example}\noindent\textbf{#1.} }{}

\newenvironment{Definition}[1][\bf Definition \thesection.\arabic{Definition}]{
        
        \refstepcounter{Definition}\noindent\textbf{#1.} }{}
\newenvironment{Proof}[1][\rm\it Proof]{%
       \noindent\textit{#1.\ } }{\nopagebreak\nolinebreak\quad\nolinebreak$\Box$\skp}

\newcounter{tmpc}

\newenvironment{List}{%
\setcounter{tmpc}{0}
\begin{list}{{\rm (\alph{tmpc})}}{\usecounter{tmpc}
\setlength{\leftmargin}{16pt}
\setlength{\rightmargin}{0cm}
\setlength{\itemsep}{2.5pt}
\setlength{\topsep}{5pt}
\setlength{\labelsep}{2pt}
\setlength{\labelwidth}{13pt}
\setlength{\listparindent}{18pt}}
}{\end{list}}

\begin{document}

%
%
%
%

\title{Jacobi Identity for Vertex Algebras in Higher Dimensions}

\author{Bojko Bakalov \,}


\address{Department of Mathematics, North Carolina State University,
Box 8205, Raleigh, NC 27695, USA}

\email{bojko\_bakalov@ncsu.edu}

\author{\, Nikolay M. Nikolov}

\address{Institute for Nuclear Research and Nuclear Energy,
Tsarigradsko Chaussee 72, BG--1784 Sofia, Bulgaria}
\curraddr{Institut f\"ur Theoretische Physik, Universit\"at G\"ottingen,
Frie\-drich--Hund--Platz 1, D--37077 G\"ottingen, Germany}

\email{mitov@inrne.bas.bg}

\date{January 6, 2006; Revised February 6, 2006}



\vspace{-12pt}

\begin{abstract}
Vertex algebras in higher dimensions provide an algebraic framework
for investigating axiomatic quantum field theory with global 
conformal invariance. We develop further the theory of 
such vertex algebras by introducing formal calculus techniques
and investigating the notion of polylocal fields.
We derive a Jacobi identity
which together with the vacuum axiom can be taken
as an equivalent definition of vertex algebra.
\end{abstract}

\maketitle

\tableofcontents


\section{Introduction}\label{se0}
\setcntrs

Two--dimensional conformal field theory is important in physics
as providing models of quantum field theory (QFT).
It also plays a role in other areas of mathematical physics
as well as in statistical physics and condensed matter physics.
A \emph{vertex algebra} is essentially the same as a \emph{chiral algebra}
in conformal field theory (see \cite{BPZ,G} and the book \cite{DMS}).
In more details, the field content in two--dimensional conformal field theories
splits into two chiral parts consisting of
fields that depend separately on one of the two light--cone variables.
Observable chiral fields have commutators supported
on the diagonal, i.e., vanishing for non-coinciding arguments.
It turns out that chiral fields form a purely algebraic structure
under their \textit{operator product expansion}, 
which is called a vertex algebra. 
Axiomatically, the notion of vertex algebra was first introduced by 
R.~E.~Borcherds \cite{B1}.
Vertex algebras arose naturally in the representation theory
of infinite--dimensional Lie algebras and in the construction
of the ``moonshine module'' for the Monster finite simple group
\cite{B1,FLM}.
Now the theory of vertex algebras is a rapidly
developing area of mathematics (see the books \cite{FLM,K,FB,LL}).
``Multi--dimensional'' generalizations of vertex algebras were considered in 
\cite{B2,KO,L2,N05}. 

The vertex algebras introduced in \cite{N05} 
for higher space--time dimension
arose naturally within a one--to--one correspondence with 
axiomatic QFT models 
satisfying the additional symmetry condition of 
\emph{global conformal invariance} (GCI).
The incorporation of GCI within the framework of axiomatic 
QFT, together with the problem of finding (nonperturbatively) 
such models in higher dimensions, has been studied previously
in e.g.\ \cite{NT,NST,NT05,NRT} 
(see also the groundbreaking early work \cite{T}).
In this way constructing models of higher--dimensional QFT with GCI becomes 
a purely algebraic problem.
Let us point out that even for general QFT (without GCI) there are not any
known models that satisfy the Wightman axioms in 
space--time dimension greater or equal to four, which
cannot be realized by free or generalized free (Heisenberg) fields.
In fact, this problem has remained
open for more than fifty years.

On the other hand, even in dimension one (i.e., 
in chiral conformal field theory), vertex algebras 
are far from full classification and are quite intricate in general.
A different algebraic structure known as a
\textit{vertex Lie algebra} has been introduced by
V.~G.~Kac \cite{K} (see also \cite{P,FB,DLM})
(it is also called a ``conformal algebra'' 
but we will not use this terminology here since it can be confused 
with the usual conformal Lie algebra in higher dimensions).
This is the structure formed by the commutators of fields, i.e.,
by the singular part of their operator product expansion.
Thus the relationship between vertex Lie algebras and 
vertex algebras is somewhat similar to the relationship between
Lie algebras and associative algebras.
It turns out that this new algebraic structure is more tractable and,
in particular, classification results for vertex Lie (super)algebras 
can be obtained \cite{DK,FK} (see also \cite{K,K2}).
The theory of vertex Lie algebras was further developed in e.g.\
\cite{CK,BKV}, and a ``multi--dimensional'' generalization
was considered in \cite{BDK}.

In the present paper we initiate an investigation
of the notion of vertex Lie algebra for the
vertex algebras in higher dimensions of \cite{N05}.
In dimension one, the main axiom for vertex Lie algebras is the so-called
\emph{Jacobi identity}, which is related to the Jacobi
identity of \cite{FLM} for vertex algebras
(and the Borcherds identity of \cite{K}).
Recall that in dimension one vertex algebras can be defined
in terms of the Jacobi identity (see \cite{FLM,FHL,K,LL}).
The main result of the present paper is a generalization of this 
Jacobi identity to higher dimensions. In addition,
we show that together with the vacuum axiom this identity can be taken as a
definition of vertex algebra equivalent to the definition of~\cite{N05}.

As in dimension one, we derive our Jacobi identity from certain 
``commutativity'' and ``associativity'' identities 
(cf.\ \cite{FHL,L1,BK,LL}). However, in the one--dimensional
case the Jacobi identity can be simplified so that it does not involve 
external sufficiently large parameters. This is no longer the case
in higher dimensions, and our Jacobi identity entails the same
degree of difficulty as the ``associativity'' identity.
The main difference with the one--dimensional case is that 
now the singular part of the operator product expansion
contains infinitely many terms.
Nevertheless, it follows from our Jacobi identity that
the singular modes close an algebraic structure under the commutator,
which would be the higher--dimensional analog of vertex Lie algebra.

The paper is organized as follows. 
The next two sections are devoted to an important technical preparation,
which can be useful not only for this work but also for future 
investigations of vertex algebras in higher dimensions.
This includes an introduction of several spaces of \emph{formal series} 
in Sect.~\ref{se1} and a higher--dimensional \emph{residue functional}
in Sect.~\ref{se3r}
(additional material is contained in Appendix~\ref{se2.5n1}).
In Sect.~\ref{se3} we recall the notions of fields, locality
and operator product expansion in higher dimensions,
mainly following \cite{N05},
but extending our considerations also to \emph{polylocal fields}.
Our main result, the \emph{Jacobi identity}, 
is contained in Sect.~\ref{se4}, together with several integral versions
and a \emph{commutator formula}.
Concluding remarks are presented in Sect.~\ref{conclusions}.

\section{Spaces of Formal Series}\label{se1}
\setcntrs

In this section we introduce various spaces of formal series,
which will be used throughout the paper. In particular, we define
the notion of a formal distribution, and we discuss formal series
expansions.

\subsection{Notation}\label{se1.1n3}\bsec
In this subsection we fix some notation to be used throughout the paper,
mostly following the notation of \cite{N05}.
We fix a positive integer $D$,
which will play the role of \textit{space--time dimension},
and we denote by $\z$, $\z_i$, $\w$, etc.,
$D$-component variables:
\begin{equation}\label{zziw}
\z = \left( z^1,\dots,z^D \right) \,, \quad
\z_i = \left( z_i^1,\dots,z_i^D \right) \,, \quad
\w = \left( w^1,\dots,w^D \right) \,.
\end{equation}
We will denote by $\z_{ij}$ the difference
\begin{equation}\label{zij}
\z_{ij} := \z_i-\z_j
= \left( z_i^1-z_j^1,\dots,z_i^D-z_j^D \right) \,,
\end{equation}
and not a new variable. We introduce the standard scalar product:
\begin{equation}\label{z1z2}
\z_1 \spr \z_2 := \sum_{\al=1}^D z_1^{\al} z_2^{\al}
\, , \qquad
\z^2 := \z \spr \z \, .
\end{equation}
Note that $\z^2$ stands for the Euclidean square of the vector $\z$,
while $z^2$ is its second component.

All vector spaces considered will be over the field $\CC$
of complex numbers. For a vector space $V$,
let $V[\z]$ (respectively, $V \llbracket \z \rrbracket$)
be the space of polynomials (respectively, formal power series)
in $\z$ with coefficients in $V$.
For a $1$-component variable $\varrho$, we denote by
$V \llbracket \varrho,\varrho^{-1} \rrbracket$
the space of formal power series
in $\varrho$ and $\varrho^{-1}$ with coefficients in $V$,
and by
$V \llbracket \varrho \rrbracket_\varrho
\equiv
V \llbracket \varrho \rrbracket [\varrho^{-1}]$
the space of Laurent series.

Note that $V[\z]$ is a $\CC[\z]$-module and
$V \llbracket \z \rrbracket$ is a $\CC \llbracket \z \rrbracket$-module.
We denote by $\CC[\z]_{\z^2}$
(respectively, $\CC \llbracket \z \rrbracket_{\z^2}$)
the localization of $\CC [\z]$ (respectively, $\CC \llbracket \z \rrbracket$)
with respect to the multiplicative system
$\{(\z^2)^k\}_{k=1,2,\dots}$.
Let $V [\z]_{\z^2}$ and $V \llbracket \z \rrbracket_{\z^2}$ be
the localizations of the corresponding modules.
Then $V[\z]_{\z^2}$ (respectively, $V \llbracket\z\rrbracket_{\z^2}$)
is a module over $\CC[\z]_{\z^2}$
(respectively, $\CC \llbracket\z\rrbracket_{\z^2}$).

We introduce the formal derivatives on $V [\z]$ and
$V \llbracket \z \rrbracket$:
\begin{equation}\label{dz}
\Di_{\z} := \left( \di_{z^1},\dots,\d_{z^D} \right) \,,
\qquad \di_{z^\al} := \frac{\di}{\di z^\al} \,,
\end{equation}
and the Euler and Laplace operators:
\begin{equation}\label{deula}
\z \spr \Di_{\z}  \, = \, \sum_{\al \, = \, 1}^D
z^\al \, \di_{z^\al}
\,, \qquad
\Di_{\z}^2 \, = \, \sum_{\al \, = \, 1}^D
(\di_{z^\al})^2 \,.
\end{equation}
Then a polynomial $f(\z) \in V[\z]$ is 
\textbf{homogeneous} of degree $m$ iff 
$(\z \spr \Di_{\z} - m) f(\z) = 0$; 
it is \textbf{harmonic} iff
$\Di_{\z}^2 f(\z) = 0$.
We denote by $V[\z]^\harm$ (respectively, $V\llbracket\z\rrbracket^\harm$)
the spaces of harmonic polynomials (respectively, formal power series).
Note that
a formal power series is harmonic if and only if each of its
homogeneous components is a harmonic polynomial.

Finally, we denote by
$\NNN$ the set of non-negative integers,
and by $\NN$ the set of positive integers.
The notation $N\gg0$ means that $N>0$ is sufficiently large.

\subsection{Harmonic Decomposition}\label{se1.1}\bsec
The classical harmonic decomposition is the fact that every
polynomial of $\z\in\CC^D$ can be divided by $\z^2$ with a unique
harmonic remainder. One can view this in a more abstract way
as follows.
Let $V$ be an arbitrary vector space. It is easy to see that
the linear operators $\Di_\z^2$, $\z^2$ and $\z\spr\Di_\z + D/2$
generate a representation of $\sl_2$ on $V[\z]$, namely
\beq\label{sfs1}
[\Di_\z^2, \, \z^2] = 2 \, \z\spr\Di_\z + D \,, \quad
[\z\spr\Di_\z, \, \Di_\z^2] = -2 \, \Di_\z^2 \,, \quad
[\z\spr\Di_\z, \, \z^2] = 2 \, \z^2 \,.
\eeq
In particular, we have the following useful formula
\begin{equation}\label{ddf1}
[\Di_\z^2, \, (\z^2)^n] 
= 4n (\z^2)^{n-1} \left( n-1 + \z\spr\Di_\z + D/2 \right) \,.
\end{equation}
It follows from the representation theory of $\sl_2$ that
every homogeneous polynomial $\FF(\z) \in V[\z]$
of degree $k$ can be written uniquely in the form
\beq\label{sfs2}
\FF(\z) = \sum_{2n+m=k} \, (\z^2)^n \, h_m(\z) \,,
\eeq
where $h_m(\z)$ are harmonic homogeneous polynomials of degree $m$
(see e.g.\ \cite[Lem\-ma 1.1]{N05} for a direct proof).
There is a similar harmonic decomposition
for elements $\FF(\z)$ of the localized space
$V[\z]_{\z^2}$; the only difference is that we allow
$k$ and $n$ to be negative (and the sum is still finite).

If we allow infinite sums, the largest space that
we get is the space $V\llbracket \z, 1/\z^2 \rrbracket$ of formal series
(see \cite{N05})
\begin{equation}\label{fd10}
\FF(\z) = \sum_{n\in\ZZ} \, \sum_{m=0}^\infty \, \sum_{\si=1}^{\har_m} \,
\FF_{\{n,m,\si\}} \, (\z^2)^{n} \, h_{m,\si}(\z) 
\,, \qquad\quad \FF_{\{n,m,\si\}} \in V \,.
\end{equation}
Here $\{ h_{m,\si}(\z) \}_{\si=1,\dots,\har_m}$
is a basis of the space of harmonic homogeneous polynomials of degree $m$
and 
\begin{equation}\label{harm}
\har_m = {m+D-1 \choose D-1} - {m+D-3 \choose D-1} \, .
\end{equation}
Note that the localized space $V \llbracket \z \rrbracket_{\z^2}$
introduced in Sect.~\ref{se1.1n3} can be naturally embedded 
in $V\llbracket \z, 1/\z^2 \rrbracket$ as the set of elements
\eqref{fd10} for which the sum over $n$ is bounded from below.

Elements \eqref{fd10} are called $V$-valued \textbf{formal distributions},
and the coefficients 
$\FF_{\{n,m,\si\}}$ in expansion \eqref{fd10}
are called \textbf{modes} of $\FF(\z)$.
The space $V \llbracket \z, 1/\z^2 \rrbracket$ has a natural structure
of a \textbf{differential module} over the algebra $\CC[z]_{\z^2}$
(see \cite[Sect.~1]{N05}), 
i.e., a $\CC[z]_{\z^2}$--module equipped with a compatible
action of $\CC[\Di_\z]$
so that the Leibniz rule for differentiation is satisfied.
We will review and generalize this in the next subsection.


\subsection{Differential Module Structure}\label{se1.1a}\bsec
In this subsection we will describe the algebraic structure of
the space $V \llbracket \z, 1/\z^2 \rrbracket$ in a way
that can be generalized 
to more general vector spaces of ``formal functions.''

%
%
As a consequence of harmonic decomposition \eqref{sfs2},
the vector space $V[\z]$ is naturally isomorphic to 
$V[\varrho] \, [\z]^\harm$, the vector spaces of harmonic
polynomials with coefficients polynomials in a $1$-component
variable $\varrho \equiv \z^2$. Similarly, we have
$V[\z]_{\z^2} \cong V[\varrho, \varrho^{-1}] \, [\z]^\harm$
and
\beq\label{sfs3}
V \llbracket \z \rrbracket_{\z^2} 
\cong V \llbracket \varrho \rrbracket_\varrho \,
\llbracket \z \rrbracket^\harm \,, \quad
V\llbracket \z, 1/\z^2 \rrbracket \cong 
V\llbracket \varrho, \varrho^{-1}\rrbracket \,
\llbracket \z\rrbracket^\harm \,,
\eeq
using the notation of Sect.~\ref{se1.1n3} and \ref{se1.1}.

We will now describe the differential module structures of
the spaces $V[\z]$ and $V[\z]_{\z^2}$ 
(over the algebras $\CC[\z]$ and $\CC[\z]_{\z^2}$, respectively)
in a way that is applicable to
the space $V\llbracket \z, 1/\z^2 \rrbracket$ and is suitable
for generalization.
Set $R = V[\varrho]$ or $V[\varrho, \varrho^{-1}]$, respectively.
Let $f(\varrho) \in R$ and let $h(\z) \in R[\z]^\harm$ 
be a harmonic homogeneous polynomial of degree $m$.
Then for each $\al=1,\dots,D$, the polynomials
$\di_{z^\al} h(\z)$ and
\begin{equation}\label{hd7}
(A_\al h)(\z) := 
z^\al \, h(\z) - (D + 2m-2)^{-1} \, \z^2 \,\, \di_{z^\al} h(\z)
\end{equation}
are harmonic as well. Indeed, for $(A_\al h)(\z)$ this follows
from \eqref{ddf1}. Note that the right--hand side of \eqref{hd7}
is well defined for $m=0$ since in this case $\di_{z^\al} h(\z) = 0$.
We deduce that the action of $z^\al$ and $\di_{z^\al}$ on 
$R[\z]^\harm$ is given by the formulas:
\begin{align}
\label{hd8}
z^\al \bigl( f(\varrho) \, h(\z) \bigr)
&= f(\varrho) \, (A_\al h) (\z)
+ (D + 2m-2)^{-1} \, \varrho f(\varrho)  \, (\di_{z^\al} h)(\z)
\,, \\
\label{hd9}
\di_{z^\al} \bigl( f(\varrho) \, h(\z) \bigr)
&= f(\varrho) \, (\di_{z^\al} h)(\z) 
+ 2 \, z^\al \, \bigl( f'(\varrho) \, h(\z) \bigr) 
\,,
\end{align}
where $\varrho \equiv \z^2$ and 
$f'(\varrho)$ denotes the derivative $d f(\varrho) / d \varrho$,
and in the right--hand side
of \eqref{hd9} one has to apply \eqref{hd8} in order to get a result
in $R[\z]^\harm$.

Now we observe that by linearity
Eqs.~\eqref{hd8}, \eqref{hd9} give rise to
a well-defined action of $z^\al$ and $\di_{z^\al}$ 
on the space of harmonic formal power series
$R\llbracket \z\rrbracket^\harm$.
This follows from the fact that the linear operators
$A_\al$ and $\di_{z^\al}$ on $R\llbracket \z\rrbracket^\harm$
are \emph{graded} (of degree $+1$ and $-1$, respectively)
with respect to the polynomial degree in $\z$.
We also notice that the right--hand sides of Eqs.~\eqref{hd8}, \eqref{hd9}
involve only the differential $\CC[\varrho]$--module structure of $R$.
These observations are summarized in the following statement.

\skp

\begin{Proposition}\label{pr1.1n1}
Let\/ $R$ be a differential module over\/ $\CC[\varrho]$
with a derivation\/ $f\mapsto f'$.
Then Eqs.~\eqref{hd8}, \eqref{hd9} define
on\/ $R[\z]^\harm$ and\/ $R\llbracket\z\rrbracket^\harm$ 
structures of differential
modules over\/ $\CC [\z]$ with derivations\/ $\di_{z^{\al}}$.
If\/ $R$ is a differential\/ $\CC[\varrho,\varrho^{-1}]$--module,
then\/ $R[\z]^\harm$ and\/ $R\llbracket\z\rrbracket^\harm$ 
are differential\/ $\CC [\z]_{\z^2}$--modules.
\end{Proposition}

\skp

\begin{Proof}
We have to check that for every $\FF (\z) \in R[\z]^\harm$ or
$R\llbracket\z\rrbracket^\harm$, the following relations 
are satisfied:
\beqa
&
z^{\al} \hspace{1pt} z^{\be} \hspace{1pt} \FF (\z)
\hspace{1pt} = \hspace{1pt} 
z^{\be}  \hspace{1pt} z^{\al} \hspace{1pt} \FF (\z)
\,, \qquad
\mathop{\sum}\limits_{\al \, = \, 1}^D \hspace{1pt}
(z^{\al})^2 \hspace{1pt} \FF (\z)
\hspace{1pt} = \hspace{1pt} 
\z^2 \hspace{1pt} \FF (\z)
\,,
& \nn &
\di_{z^{\al}} \hspace{1pt} \di_{z^{\be}} \hspace{1pt} \FF (\z)
\hspace{1pt} = \hspace{1pt}
\di_{z^{\be}} \hspace{1pt} \di_{z^{\al}} \hspace{1pt} \FF (\z)
\,, \qquad
\di_{z^{\al}} \hspace{1pt} (z^{\be} \hspace{1pt} \FF (\z))
\hspace{1pt} = \hspace{1pt}
z^{\be} \hspace{1pt} \di_{z^{\al}} \hspace{1pt} \FF (\z)
+ \delta_{\al}^{\be} \hspace{1pt} \FF (\z)
\,,
& \nn &
\di_{z^{\al}} \hspace{1pt} ((\z^2)^{-1} \hspace{1pt} \FF (\z))
\hspace{1pt} = \hspace{1pt}
(\z^2)^{-1} \hspace{1pt} \di_{z^{\al}} \hspace{1pt} \FF (\z)
- 2 \hspace{1pt} (\z^2)^{-2} \hspace{1pt} z^{\al} \hspace{1pt} \FF (\z)
\,.
& \nonumber \raisebox{13pt}{}
\eeqa
This can be verified by a straightforward computation, or can be deduced
from the fact that these relations hold for 
$R=V[\varrho]$ and $V[\varrho, \varrho^{-1}]$.\end{Proof}

\skp

In particular, using isomorphism \eqref{sfs3} and
taking $R = V\llbracket \varrho, \varrho^{-1}\rrbracket$
in the above proposition, we obtain a structure of a 
differential $\CC [\z]_{\z^2}$--module
on the space 
$V\llbracket \z,$ $1/\z^2 \rrbracket$ 
of formal distributions
(cf.\ \cite{N05}).
Note that $V\llbracket \z \rrbracket$ is also a 
$\CC\llbracket \z \rrbracket$--module,
and $V\llbracket \z \rrbracket_{\z^2}$ is a 
$\CC\llbracket \z \rrbracket_{\z^2}$--module.
However, 
$V\llbracket \z, 1/\z^2 \rrbracket$ is \emph{not}
a $\CC\llbracket \z \rrbracket$--module because
$V\llbracket \varrho,$ $\varrho^{-1}\rrbracket$
is not a $\CC\llbracket \varrho \rrbracket$--module.
Although obvious, the next remark plays an important role
in the theory.

\skp

\begin{Remark}\label{rhd}
The action of $\CC\llbracket\z\rrbracket_{\z^2}$ 
on $V\llbracket\z\rrbracket_{\z^2}$ 
does not have zero divisors. In other words,
if $p(\z)\, \FF(\z) = 0$ for $p(\z) \in \CC\llbracket\z\rrbracket_{\z^2}$,
$\FF(\z) \in V\llbracket\z\rrbracket_{\z^2}$, 
then either $p(\z)=0$ or $\FF(\z)=0$.
Note that this is not the case for the $\CC [\z]_{\z^2}$--module
$V\llbracket \z, 1/\z^2 \rrbracket$ (see \cite[Example~1.1]{N05}).
\end{Remark}

\subsection{Generalized Formal Distributions}\label{se1.1b}\bsec
As another application of \ref{pr1.1n1}, we will define spaces
of formal distributions that involve non-integral powers of $\z^2$.
We will take $R$ to be
a space of formal series in real powers of $\varrho$,
\beq\label{eq1.12n1}
R = V \llbracket \varrho^{\Gamma} \rrbracket
\, := \,
\left\{\raisebox{12pt}{\hspace{-2pt}}\right.
f(\varrho) =
\mathop{\sum}\limits_{\gamma \, \in \, \Gamma} \,
f_{\gamma} \,
\varrho^{\gamma}
\; \Big| \;
f_{\gamma} \in V
\left.\raisebox{12pt}{\hspace{-2pt}}\right\}
\,,
\eeq
where $\Gamma$ is a \textbf{$\ZZ$--invariant} subset of $\RR$, i.e.,
such that
\beq\label{eq2.13n3}
\Gamma + \ZZ \ := \left\{ \gamma + m \; | \;
\gamma \in \Gamma, \, m \in \ZZ \right\}
\subseteq \Gamma \,.
\eeq
Then $R$ is a differential $\CC[\varrho,\varrho^{-1}]$--module
(with $(\varrho^\gamma)' = \gamma\varrho^{\gamma-1}$),
and $R\llbracket\z\rrbracket^\harm$ is a 
differential $\CC [\z]_{\z^2}$--module, which we denote
as $V \llbracket \z,(\z^2)^{\Gamma} \rrbracket$.
We can view the elements of 
$V \llbracket \z,(\z^2)^{\Gamma} \rrbracket$ as infinite series
(cf.\ \eqref{fd10}, \eqref{sfs3})
\begin{equation}\label{fd11}
\FF(\z) = \sum_{\ga\in\Ga} \, \sum_{m=0}^\infty \, \sum_{\si=1}^{\har_m} \,
\FF_{\{\ga,m,\si\}} \, (\z^2)^\ga \, h_{m,\si}(\z) 
\,, \qquad\quad \FF_{\{n,m,\si\}} \in V \,.
\end{equation}

For $\Ga=\ZZ$, the above-defined module
$V \llbracket \z,(\z^2)^{\ZZ} \rrbracket$ coincides
with $V \llbracket \z,1/\z^2 \rrbracket$.
We will call elements \eqref{fd11} 
(generalized) \textbf{formal distributions}.
Note that this construction works also for subsets
$\Ga\subseteq\CC$ but we will restrict our considerations to $\RR$.
We define inductively the vector spaces 
\beq\label{eq2.16n3}
\begin{split}
V \llbracket \z_1, (\z_1^2)^{\Gamma_1} & ; \dots; 
\z_n \, ,(\z_n^2)^{\Gamma_n} \rrbracket
\\
&:= 
\Bigl( V \llbracket \z_1, (\z_1^2)^{\Gamma_1} ; \dots; 
\z_{n-1},(\z_{n-1}^2)^{\Gamma_{n-1}} \rrbracket \Bigr) 
\llbracket \z_n \, , (\z_n^2)^{\Gamma_n} \rrbracket
\end{split}
\eeq
of formal distributions in several $D$--dimensional vector variables 
$\z_1,\dots,\z_n$.
When all $\Gamma_i=\ZZ$, we will denote the space \eqref{eq2.16n3} as
$V \llbracket \z_1, 1/\z_1^2 \, ; \dots; \z_n,1/\z_n^2 \rrbracket$.
%

\skp

\begin{Remark}\label{rm-harm2}
Choosing instead $R = V[\varrho^\Gamma]$ in the above construction, 
we obtain a differential $\CC [\z]_{\z^2}$--module 
$R[\z]^\harm$ denoted as $V[ \z,(\z^2)^{\Gamma} ]$;
it consists of all finite sums of the form \eqref{fd11}.
\end{Remark}

\skp

\begin{Remark}\label{rm-harm}
Note that Eq.~\eqref{ddf1} is valid for $n\in\Gamma$, and it implies
\begin{equation}\label{ddf2}
\Di_\z^2 \bigl( (\z^2)^\ga \, h_{m,\si}(\z) \bigr)
= 4\ga \Bigl( \ga+ m + \frac{D}{2} - 1 \Bigr)
(\z^2)^{\ga-1} \, h_{m,\si}(\z) 
\,.
\end{equation}
In particular, $(\z^2)^\ga \, h_{m,\si}(\z)$ is harmonic if
and only if $\ga=0$ or $\ga=-\frac{D}{2}+1-m$.
\end{Remark}


\skp

\begin{Example}\label{rm2.1n5}
Let us consider the case $D = 1$.
Then $\z = z$ is a $1$-component variable and the harmonic polynomials
are just the affine polynomials $a+bz$.
Thus, every element of
$V \llbracket z,(z^2)^{\Gamma} \rrbracket$ has the form
\beq\label{eq2.14n4}
\FF (z) =
\mathop{\sum}\limits_{\gamma \, \in \, \Gamma} \,
(z^2)^{\gamma} \, (a_{\gamma} + b_{\gamma} \, z) \,, \qquad
a_{\gamma},b_{\gamma} \in V \,.
\eeq
It is easy to see that when $\Gamma$ is an additive subgroup of $\RR$
containing $\frac12\ZZ$, we have the following
direct sum of differential $\CC [z]_{z^2}$--modules:
\beq\label{eq2.15n4}
V \llbracket z,(z^2)^{\Gamma} \rrbracket
\, = \,
(1 \hspace{-1pt} + \hspace{-1pt} (z^2)^{-1/2} \, z) \,
V \llbracket z,(z^2)^{\Gamma} \rrbracket
\, \oplus \,
(1 \hspace{-1pt} - \hspace{-1pt} (z^2)^{-1/2} \, z) \,
V \llbracket z,(z^2)^{\Gamma} \rrbracket \,.
\eeq
Note that here $(z^2)^{-1/2} \, z$ is viewed as an element of
$V \llbracket z,(z^2)^{\Gamma} \rrbracket$, and is not equal to~$1$.
\end{Example}

\subsection{$\Gamma$--Localization}\label{se1.1g}\bsec
The vector space of localized formal power series
$V \llbracket \z\rrbracket_{\z^2}$ can be embedded in
$V \llbracket \z,$ $1/\z^2 \rrbracket$ as the subspace
of all elements \eqref{fd10} for which the sum over $n$
is bounded from below. This space is a (differential) module over the 
localized algebra $\CC \llbracket \z\rrbracket_{\z^2}$.
We will later need the following generalizations of these spaces.

Let $\Ga_1,\dots,\Ga_s$ be 
additive subgroups of $\RR$ 
containing $\ZZ$; then each $\Gamma_i$ is automatically $\ZZ$-invariant
(see \eqref{eq2.13n3}).
Let $A$ be a commutative associative algebra, and let
$f_1,\dots,f_s$ be some fixed elements of $A$.
Then we define
\beq\label{eq1.13n1}
A_{f_1^{\Ga_1} \com\dotsm\com f_s^{\Ga_s}}
\, := \,
\bigl(
\CC[\Ga_1] \otimes_\CC \dotsm \otimes_\CC \CC[\Ga_s] \otimes_\CC A
\bigr) \big/ J \,,
\eeq
where $\CC[\Ga_i]$ is the group algebra of $\Ga_i$ (with elements denoted as
$e^\ga$) and $J$ is the ideal generated by elements
of the form
\beq\label{eq1.14n1}
e^{\gamma_1'} \otimes \dotsm \otimes e^{\gamma_s'} \otimes g
-
e^{\gamma_1''} \otimes \dotsm \otimes e^{\gamma_s''} \otimes g
\eeq
for which there exist $\gamma_i \in \Gamma_i$ such that
$\gamma_i+\gamma_i',\gamma_i+\gamma_i'' \in \NNN$ $(i=1,\dots,s)$ and
\beq\label{eq1.15n1}
f_1^{\gamma_1+\gamma_1'} \dotsm \, f_s^{\gamma_s+\gamma_s'} \, g
=
f_1^{\gamma_1+\gamma_1''} \dotsm \, f_s^{\gamma_1+\gamma_s''} \, g
\eeq
in the algebra $A$.
The so-defined commutative associative algebra 
$A_{f_1^{\Ga_1} \com\dotsm\com f_s^{\Ga_s}}$ will be called
the $(\Ga_1,\dots,\Ga_s)$--\textbf{localization} of $A$ with respect to 
$f_1,\dots,f_s$.
When all groups $\Ga_i$ are equal to $\Ga$, we will just call it
$\Ga$--localization.
The image in $A_{f_1^{\Ga_1} \com\dotsm\com f_s^{\Ga_s}}$ of an element
$e^{\gamma_1} \otimes \dotsm \otimes e^{\gamma_s} \otimes g$
will be denoted as
$f_1^{\gamma_1} \dotsm \, f_s^{\gamma_s} \, g$.

Obviously, the $\ZZ$--localization of $A$ 
with respect to $f_1,\dots,f_s$ coincides with the localized
algebra $A_{f_1 \dotsm f_s}$. In this case the localization
with respect to a set $\{f_1,\dots,f_s\}$ is naturally isomorphic to the
localization with respect to the product $f_1 \dotsm f_s$.
For general $\Gamma$ this is not true.

If $M$ is an $A$--module, then in the same way
one defines the localization
$M_{f_1^{\Ga_1} \com\dotsm\com f_s^{\Ga_s}}$ as an
$A_{f_1^{\Ga_1} \com\dotsm\com f_s^{\Ga_s}}$--module.
In addition, if $M$ is a differential $A$--module,
then 
the localization
$M_{f_1^{\Ga_1} \com\dotsm\com f_s^{\Ga_s}}$ is a differential
$A_{f_1^{\Ga_1} \com\dotsm\com f_s^{\Ga_s}}$--module
with the same set of derivations.
Indeed, every derivation of $A$ extends
by the Leibniz rule to $A_{f_1^{\Ga_1} \com\dotsm\com f_s^{\Ga_s}}$
since $J$ is an invariant subspace.
We note also that if $M$ has no zero divisors
the same is true for $M_{f_1^{\Ga_1} \com\dotsm\com f_s^{\Ga_s}}$
(cf.\ \ref{rhd}).

As a special case of the above construction, 
we get a differential $\CC \llbracket\z\rrbracket_{(\z^2)^\Gamma}$--module
$V \llbracket\z\rrbracket_{(\z^2)^\Gamma}$. 
The latter can be identified with the $\CC [\z]_{\z^2}$--submodule of
$V \llbracket\z,$ $(\z^2)^{\Gamma}\rrbracket$ 
consisting of all elements \eqref{fd11} for which the sum over $\gamma$
is over the union of \emph{finitely} many 
sets of the form $\{ \ga_i + \NNN \}$.

\subsection{Formal Expansions}\label{se1.2}\bsec
Let $\Gamma$ be 
an
additive subgroup of $\RR$ containing $\ZZ$,
and let $V$ be a vector space.
Recall the $\Gamma$--localizations defined in Sect.~\ref{se1.1g}
and the spaces of formal series defined by \eqref{fd11}, \eqref{eq2.16n3}.
There are obvious embeddings
\begin{equation}\label{fex9}
\CC\llbracket\z,\w\rrbracket_{(\z^2)^\Gamma \com (\w^2)^\Gamma} 
\subsetneqq
\CC\llbracket\z\rrbracket_{(\z^2)^\Gamma} \,
\llbracket\w\rrbracket_{(\w^2)^\Gamma} 
\subsetneqq
\CC\llbracket\z, (\z^2)^\Gamma ; \w, (\w^2)^\Gamma \rrbracket \,.
\end{equation}
%
%
We emphasize that the spaces
$\CC\llbracket\z\rrbracket_{\z^2}
\llbracket\w\rrbracket_{\w^2}$
and 
$\CC\llbracket\w\rrbracket_{\w^2}
\llbracket\z\rrbracket_{\z^2}$
are different, because elements of the former space
have only finitely many negative powers of $\w^2$
but possibly infinitely many negative powers of $\z^2$.
Note also that the first two spaces in \eqref{fex9} are rings, and hence
the second one is a module over the first.

For $\gamma \in \Gamma$, we define
$\iota_{\z,\w} \hspace{1pt} ((\z - \w)^2)^{\gamma}$ as the
Taylor expansion of $((\z - \w)^2)^{\gamma}$ in $\w$ around $0$, to wit
\beq\label{eq2.21n2}
\iota_{\z,\w}
\bigl( (\z-\w)^2 \bigr)^{\gamma}
\hspace{-1pt} :=
e^{-\w \spr \Di_{\z}} \, (\z^2)^{\gamma}
 = \! \mathop{\sum}\limits_{k \, = \, 0}^{\infty} \,
\frac{(-\w \spr \Di_{\z})^k}{k!} \, (\z^2)^{\gamma}
\, \in \, \CC \llbracket \z \rrbracket_{(\z^2)^\Gamma} \,
\llbracket \w \rrbracket \,. \;
\eeq
One defines 
$\iota_{\z,\w} \hspace{1pt} ((\z + \w)^2)^{\gamma}$ 
in the same way, while
$\iota_{\w,\z} \hspace{1pt} ((\z - \w)^2)^{\gamma}$
is defined using the identifications
\beq\label{zwgwzg}
\bigl( (\z - \w)^2 \bigr)^{\gamma} \equiv
\bigl( (\w - \z)^2 \bigr)^{\gamma} \,,
\qquad \bigl( (-\z)^2 \bigr)^\ga \equiv (\z^2)^\ga \,.
\eeq

More generally, for
$\psi (\z,\w) \in V \llbracket \z,\w 
\rrbracket_{(\z^2)^\Gamma \com (\w^2)^\Gamma}$
we set
\beq\label{eq2.22n2}
\iota_{\z,\w}
\left(\raisebox{10pt}{\hspace{-3.5pt}}\right.
\hspace{1pt} \psi (\z,\w)
\left(\raisebox{9pt}{\hspace{-2.5pt}}\right.
(\z-\w)^2
\left.\raisebox{9pt}{\hspace{-2.5pt}}\right)^{\gamma}
\left.\raisebox{10pt}{\hspace{-3pt}}\right)
\, :=  \,
\psi (\z,\w) \,
\iota_{\z,\w}
\left(\raisebox{9pt}{\hspace{-2.5pt}}\right.
(\z-\w)^2
\left.\raisebox{9pt}{\hspace{-2.5pt}}\right)^{\gamma} \,,
\eeq
thus obtaining a
$\CC \llbracket \z,\w 
\rrbracket_{(\z^2)^\Gamma \com (\w^2)^\Gamma}$--linear map
\beq\label{eq2.23n2}
\iota_{\z,\w} \colon
V \llbracket \z,\w \rrbracket_{(\z^2)^\Gamma \com (\w^2)^\Gamma
\com ((\z-\w)^2)^\Gamma}
\, \longrightarrow \,
V \llbracket \z \rrbracket_{(\z^2)^\Gamma} \,
\llbracket \w \rrbracket_{(\w^2)^\Gamma} \,.
\eeq
Note that the map $\iota_{\z,\w}$ commutes with all partial derivatives
$\di_{z^{\al}}$ and $\di_{w^{\al}}$.

We define another version of $\iota_{\z,\w}$ as follows:
\beq\label{eq2.29n3}
\begin{split}
\hiota_{\z,\w} \colon
V \llbracket \z-\w \rrbracket_{((\z-\w)^2)^\Ga} \,
\llbracket \w \rrbracket_{(\w^2)^\Ga}
\, \longrightarrow \,
& \;
V \llbracket \z \rrbracket_{(\z^2)^\Ga} \,
\llbracket \w \rrbracket_{(\w^2)^\Ga} \,,
\\ \raisebox{15pt}{}
\FF (\z-\w,\w)
\, \longmapsto \,
& \; 
e^{-\w' \spr \Di_{\z}} \FF (\z,\w)
\big|_{\w' = \w} \,,
\end{split}
\eeq
where in the left--hand side $\z\!-\!\w$ is viewed as an independent variable.
Equation~\eqref{eq2.29n3} is well defined because
$e^{-\w' \spr \Di_{\z}} \FF (\z,\w)$ $\in$
$V \llbracket \z \rrbracket_{(\z^2)^\Ga}
\llbracket \w,\w' \rrbracket_{(\w^2)^\Ga}$
(while an analogous formal expansion $\hiota_{\z,\w}$
on $V \llbracket \w \rrbracket_{(\w^2)^\Ga}
\llbracket \z-\w \rrbracket_{((\z-\w)^2)^\Ga}$
does not make sense).
Obviously, maps \eqref{eq2.23n2} and \eqref{eq2.29n3} 
agree with each other on the intersection of their domains.
The two versions of $\iota_{\z,\w}$ can be related to each other
by the following statement.

\skp

\begin{Proposition}\label{pr2.2n3}%
\hspace{0pt}{\rm (Taylor's formula.)}\hspace{-1pt}
For every $\FF (\z,\w)$ $\in$
$V \llbracket \z,$ 
$\w \rrbracket_{(\z^2)^\Ga \com (\w^2)^\Ga \com ((\z-\w)^2)^\Ga}$
we have
\beq\label{eq2.30n3}
\hiota_{\z,\w} \
\iota_{\z-\w,\w} \ \FF ((\z-\w)+\w,\w)
\, = \, \iota_{\z,\w} \ \FF (\z,\w) \,.
\eeq
\end{Proposition}

\skp

\begin{Proof}
It suffices to prove (\ref{eq2.30n3}) for $\FF (\z,\w) = (\z^2)^{\ga}$,
$\ga \in \Ga$.
Then, according to Eqs.~(\ref{eq2.21n2}) and (\ref{eq2.29n3}),
we have:
\beqa
\hspace{40pt}
\hiota_{\z,\w} \
\iota_{\z-\w,\w} \ \bigl( ((\z-\w)+\w)^2 \bigr)^\ga
=
\hiota_{\z,\w}
\bigl(
e^{\w' \spr\Di_{\z}} \ \bigl( (\z-\w)^2 \bigr)^\ga \big|_{\w' = \w}
\bigr)
&& \nn \hspace{40pt} =
e^{-\w'' \spr\Di_{\z}} \
e^{\w' \spr\Di_{\z}} \ (\z^2)^\ga \big|_{\w' = \w'' = \w}
\, = \, (\z^2)^\ga \,,
\nonumber
\eeqa
which completes the proof.\end{Proof}

\skp

We define the spaces of successively localized formal series
\beq\label{suc-loc}
V \llbracket \z_1 \rrbracket_{(\z_1^2)^{\Gamma}} \dotsm
\llbracket \z_1 \rrbracket_{(\z_n^2)^{\Gamma}}
\, := \,
\Bigl( 
V \llbracket \z_1 \rrbracket_{(\z_1^2)^{\Gamma}}
\dotsm \llbracket \z_{n-1} \rrbracket_{(\z_{n-1}^2)^{\Gamma}}
\Bigr) \llbracket \z_n \rrbracket_{(\z^2_n)^{\Gamma}} \,,
\eeq
which will be used in the sequel.
The space \eqref{suc-loc} is a module over the algebra 
$\CC \llbracket \z_1 \rrbracket_{(\z_1^2)^{\Gamma}} \dotsm
\llbracket \z_1 \rrbracket_{(\z_n^2)^{\Gamma}}$.
Again, one should keep in mind that in \eqref{suc-loc} 
one would get a different space if the variables are put
in different order.

\section{Residue Functional}\label{se3r}
\setcntrs

In this section we introduce an important linear functional
on the spaces of formal distributions, which we call the residue
functional. We discuss its fundamental properties and we prove
an analog of the Cauchy formula.
A geometric interpretation of the residue is given in
Appendix~\ref{se2.5n1}.

\subsection{Definition and Main Properties}\label{se2.3n2}\bsec
In this subsection we introduce, for an arbitrary vector space $V$,
an important linear map
\(V \llbracket \z, (\z^2)^{\RR} \rrbracket \to V\),
which will be denoted as
$\FF (\z) \mapsto \Res_{\z} \, \FF (\z)$ and will be called
the \textbf{residue}.
Observing that every element of $V \llbracket \z, (\z^2)^{\RR} \rrbracket$
can be uniquely represented as a formal series of the form
\beq\label{eq1.17n2}
\FF (\z) = \sum_{\gamma \, \in  \, \RR} \,
( \z^2 )^{\gamma} \, \FF_{\gamma} ( \z )
\qquad \text{with} \quad
\FF_{\gamma} (\z) \in V\llbracket \z \rrbracket^\harm \,,
\eeq
we define
\begin{equation}\label{eqn5}
\Res_{\z} \,
\sum_{\gamma \, \in  \, \RR} \,
( \z^2 )^{\gamma} \FF_{\gamma} ( \z )
\, := \,
\FF_{-\frac{D}2} ( \zero )
\, . \
\end{equation}

\skp

\begin{Theorem}\label{prp2n.2}
{\rm{(a)}}
The linear map~\eqref{eqn5} is \emph{$\Di_{\z}$--invariant}
in the sense that
\beq\label{eqn6}
\Res_{\z} \, \di_{z^{\al}} \FF(\z) \, = \, 0
\
\eeq
for all \(\FF(\z) \in V \llbracket \z, (\z^2)^{\RR} \rrbracket\)
and \(\al = 1,\dots,D\).

\medskip

{\rm{(b)}}
The bilinear form
\beq\label{eqn7n1}
\langle f ,\, g \rangle \, := \, \Res_{\z} \, f(\z) g(\z)
\, , \qquad
f,\, g \in \CC [\z, (\z^2)^{\RR}]
\, , \
\eeq
is \emph{nondegenerate}.

\medskip

{\rm{(c)}}
Let $h_{m} ( \z )$ and
$h'_{m'} ( \z )$ be
harmonic homogeneous polynomials of degrees $m$ and $m'$, respectively.
Then 
\beq\label{eqn8n1}
\begin{split}
\langle
( \z^2 )^{\gamma} h_{m} (\z)
& , \,
h'_{m'} (\z)
\rangle
= 0 \quad  \text{if}
\\
& m \hspace{-1pt} \neq m' \quad \text{or} \quad
2\gamma + m + m' \neq \hspace{-1pt} - D \,,
\end{split}
\eeq
and in the opposite case this coincides with the unique,
up to a multiplicative constant,
$\Oo(D)$--invariant
scalar product on the vector space of harmonic homogeneous polynomials
of degree
\(m \, ( \, =m')\) given by
\beq\label{eqn8n2}
\langle
(\z^2)^{-m-\frac{D}{2}} h_{m} (\z)
, \, h'_{m} (\z)
\rangle
\,.
\eeq
\end{Theorem}

\skp

\begin{Proof}
(a)
Let $h(\z)$ be a harmonic homogeneous polynomial of degree $m$.
{}From Eqs. \eqref{hd7}--\eqref{hd9} we deduce the harmonic decomposition
\begin{align*}
\di_{z^\al} \bigl( (\z^2)^\ga \, h(\z) \bigr)
= & \,
2\ga \, (\z^2)^{\ga-1} \, (A_\al h)(\z)
\\
& + \bigl( 1 + 2\ga ( D + 2m-2 )^{-1} \bigr) 
\, (\z^2)^\ga  \, \di_{z^\al} h(\z) 
\,.
\end{align*}
Now let us apply $\Res_{\z}$ to the right--hand side of this equation.
The first term will give zero, because $(A_\al h)(\zero) = 0$.
Similarly, the second term can give a nonzero
result only if \(m = 1\) and \(\gamma = -D/2\)
but then the coefficient vanishes.

(b, c)
Property~(\ref{eqn8n1}) 
follows from the harmonic decomposition
\[
h_{m} ( \z ) \, h'_{m'} ( \z )
\, = \,
\mathop{\sum}\limits_{n \, = \, 0}^{\min ( m,m' )}
( \z^2 )^{n} \, h''_{m+m'-2n} ( \z ) \,,
\]
where $h''_{m''}(\z)$ are uniquely determined harmonic homogeneous polynomials
of degree $m'' = |m-m'|,\dots,m+m'$.
It is known that for $m = m'$, 
the constant polynomial $h''_0 \in \CC$ defines
an $\Oo(D)$--invariant nondegenerate scalar product on the space of
harmonic polynomials of degree $m$. This proves the remaining statements.\end{Proof}

\skp

{}From now on, we will assume that the bases
$\{ h_{m,\si}(\z) \}_{\si=1,\dots,\har_m}$
of harmonic homogeneous polynomials of degree $m$
are \textit{orthonormal}, so that
\beq\label{eqn2n1}
\Res_{\z} \hspace{2pt} (\z^2)^\ga 
\hspace{2pt} h_{m,\sigma} (\z) \hspace{2pt} h_{m',\sigma'} (\z) \, = \,
\delta_{\ga,-\frac{D}2-m} \hspace{2pt} \delta_{m,m'} \hspace{2pt} 
\delta_{\sigma,\sigma'} \,,
\eeq
in accord with \ref{prp2n.2}(c).
Consequently, the modes of a formal series $\FF(\z)$
given by \eqref{fd11} can be recovered as residues:
\begin{equation}\label{fd12}
\FF_{\{\ga,m,\si\}} = \Res_{\z} \, \FF(\z) \, 
(\z^2)^{-\frac{D}2-\ga-m} \, h_{m,\si}(\z) \,.
\end{equation}
This justifies the name ``formal distributions.''

\skp

\begin{Corollary}\label{pr2.3n3}
Let\/ $\Gamma$ be a\/ $\ZZ$--invariant subset of\/ $\RR$
$($i.e., $\Gamma + \ZZ \subseteq \Gamma)$, and set
\beq\label{eq2.36n3}
\Gamma'
= - \Gamma + \frac{D}{2}
:=
\left\{\raisebox{12pt}{\hspace{-2pt}}\right.
-\gamma + \frac{D}{2} \; \Big| \; \gamma \in \Gamma
\left.\raisebox{12pt}{\hspace{-2pt}}\right\} \,.
\eeq
Then
\beq\label{eq2.37n3}
V \llbracket \z,(\z^2)^{\Gamma} \rrbracket \, \cong \,
\Hom_{\CC}
\left(\raisebox{9pt}{\hspace{-2pt}}\right.
\CC [\z,(\z^2)^{\Gamma'}] \, , \,V
\left.\raisebox{9pt}{\hspace{-2pt}}\right)
\eeq
as differential\/ $\CC [\z]_{\z^2}$--modules.
In particular,
$\CC \llbracket \z,$ $(\z^2)^{\Gamma} \rrbracket$ is the dual\/
$\CC [\z]_{\z^2}$--module of\/
$\CC [\z,(\z^2)^{\Gamma'}]$.
\end{Corollary}

\skp

Note that by the recursive definition~(\ref{eq2.16n3}) the residue
functional is defined also on formal distributions in several
$D$--dimensional variables.
For instance, we have
\beq\label{eq2.48n4}
V \llbracket \z,(\z^2)^{\RR}; \w,(\w^2)^{\RR} \rrbracket
\xrightarrow{\Res_{\w}}
V \llbracket \z,(\z^2)^{\RR} \rrbracket
\xrightarrow{\Res_{\z}}
V \,.
\eeq
Then under the natural isomorphism
\beq
V \llbracket \z, (\z^2)^{\RR}; \w, (\w^2)^{\RR} \rrbracket
\cong
V \llbracket \w, (\w^2)^{\RR}; \z, (\z^2)^{\RR} \rrbracket
\eeq
the ``Fubini theorem'' is satisfied, namely,
\beq\label{eq2.49n4}
\Res_{\z} \, \Res_{\w} \, \FF (\z,\w) \, = \,
\Res_{\w} \, \Res_{\z} \, \FF (\z,\w)
\eeq
for $\FF (\z,\w)$ $\in$
$V \llbracket \z,(\z^2)^{\RR};\w,(\w^2)^{\RR} \rrbracket$.

The following proposition describes all $\Di_{\z}$--invariant
linear functionals
$V \llbracket \z,$ $(\z^2)^{\Gamma} \rrbracket$ $\to$ $V$
(cf.\ \eqref{eqn6}).

\skp

\begin{Proposition}\label{pr2.4n5}
Let\/ $\Gamma$ be a $\ZZ$--invariant subset of\/ $\RR$,
and let\/ $\Omega \colon V \llbracket \z,$ $(\z^2)^{\Gamma} \rrbracket \to$ $V$
be a linear map
that is $\Di_{\z}$--invariant, i.e., such that
\(\Omega ( \di_{z^{\al}} \FF ) = 0\) \ for all\/
\(\FF \in V \llbracket \z, (\z^2)^{\Gamma} \rrbracket\) \ and \
\(\al = 1,\dots,D\).

\medskip

{\rm{(a)}}
If\/ 
$D \geqslant 2$,
then there exists a complex constant\/ $C$ such that\/
\(\Omega ( \FF ) = C \, \Res_{\z} \FF\) \ for all
\(\FF \in V \llbracket \z, (\z^2)^{\Gamma} \rrbracket\).
In particular, if\/ $\frac{D}{2} \notin \Gamma$ then $\Omega = 0$.

\medskip

{\rm{(b)}}
If\/ 
$D = 1$,
then there exist complex constants $C$ and $C^{\,\prime}$ such that\/
\(\Omega ( \FF ) =
C \, \Res_{z} \FF + C^{\,\prime} \, \Res_{z}' \FF\)
\ for all\/
\(\FF \in V \llbracket z, (z^2)^{\Gamma} \rrbracket\)
$(z$ is now a $1$-component variable$)$, where $\Res_{z}'$ is defined by
\beq\label{eqn5-1}
\Res_{z}' \,
\sum_{\gamma \, \in  \, \RR} \,
( z^2 )^{\gamma}
(a_{\gamma} + b_{\gamma} \, z)
\, := \,
b_{-1}
\, \
\eeq
$($see \ref{rm2.1n5}$)$.
\end{Proposition}

\skp

\begin{Proof}
The space of all $\Di_{\z}$--invariant linear maps
$\Omega \colon V \llbracket \z, (\z^2)^{\Gamma} \rrbracket \to V$
is isomorphic to
the vector space
\[
\Hom_{\CC} \bigl( V \llbracket \z, (\z^2)^{\Gamma} \rrbracket
\hspace{2pt}/\hspace{1pt}
\Di_{\z} \, V \llbracket \z, (\z^2)^{\Gamma} \rrbracket ,V \bigr) 
\,,
\]
where
\[
\Di_{\z} \, V \llbracket \z, (\z^2)^{\Gamma} \rrbracket :=
\di_{z^1} V \llbracket \z, (\z^2)^{\Gamma} \rrbracket
+ \dots +
\di_{z^D} V \llbracket \z, (\z^2)^{\Gamma} \rrbracket \,.
\]

Fix
$\gamma \in \Gamma$ and a harmonic homogeneous polynomials
$h ( \z )$ of degree $m$. We will prove that
for $D \geqslant 2$ one has
\beq\label{eq2.39n5}
( \z^2 )^{\gamma}
h ( \z )
\, \in \, \Di_{\z} V \llbracket \z,(\z^2)^{\Gamma} \rrbracket
\quad \text{if} \quad
(\gamma,m) \neq
\left(\raisebox{10pt}{\hspace{-3pt}}\right.
-\frac{D}{2},0
\left.\raisebox{10pt}{\hspace{-2pt}}\right) \,,
\eeq
while for $D = 1$ one has
\beq\label{eq2.40n5}
( \z^2 )^{\gamma}
h ( \z )
\, \in \, \Di_{\z} V \llbracket \z,(\z^2)^{\Gamma} \rrbracket
\quad \text{if} \quad
(\gamma,m) \neq
\left(\raisebox{10pt}{\hspace{-3pt}}\right.
-\frac{1}{2},0
\left.\raisebox{10pt}{\hspace{-2pt}}\right)
\ \ \text{or} \ \
(\gamma,m) \neq (-1,1) \,. \;
\eeq

Indeed, using the equalities (see \eqref{ddf2})
\[
\Di_{\z}^2 \bigl( ( \z^2 )^{\gamma+1} h ( \z ) \bigr)
\, = \,
4 ( \gamma + 1 )
\Bigl(
\gamma + m +
\frac{D}{2}
\Bigr)
( \z^2 )^{\gamma} h ( \z )
\]
and
\[
\text{\large \(\mathop{\sum}\limits_{\al \, = \, 1}^{D}\)} \,
\di_{z^{\al}}
\bigl( ( \z^2 )^{\gamma+1} \, \di_{z^{\al}} h ( \z ) \bigr)
\, = \, 2 \, m ( \gamma + 1 )
( \z^2 )^{\gamma} h ( \z ) \,,
\]
we conclude that
$( \z^2 )^{\gamma} h ( \z ) \in \Di_{\z} V
\llbracket \z,(\z^2)^{\Gamma} \rrbracket$
if $\gamma \neq -1$ and
\(( \gamma,m ) \neq
\left(\raisebox{9pt}{\hspace{-3pt}}\right.
-\frac{D}{2},0
\left.\raisebox{9pt}{\hspace{-2pt}}\right)\).
Finally, in the case $\gamma = -1$, we have
\[
\mathop{\sum}\limits_{\al \, = \, 1}^{D}
\di_{z^{\al}}
\bigl( ( \z^2 )^{-1} z^{\al} \, h (\z) \bigr) \, = \,
( D + m - 2 ) ( \z^2 )^{-1} h (\z) \,.
\]
This proves (\ref{eq2.39n5}) and (\ref{eq2.40n5}).

Now observing that by \ref{prp2n.2}(a) we have
$
( \z^2 )^{-D/2}
\notin \Di_{\z} V \llbracket \z, (\z^2)^{\Gamma} \rrbracket
$,
we complete the proof of part (a).
To prove part (b), it remains to check that $\Res_{z}'$ is
$\di_{z}$--invariant, which is straightforward.\end{Proof}

\skp

\begin{Example}\label{exresd1}
Let $D=1$; then elements of $V \llbracket z,(z^2)^{\Gamma} \rrbracket$
have the form \eqref{eq2.14n4}. In particular, for $\Gamma=\ZZ$,
we can write every element $\FF (z) \in V \llbracket z, 1/z^2 \rrbracket$
uniquely as
\beq\label{fs1d1}
\FF (z) =
\mathop{\sum}\limits_{n\in\ZZ} \,
c_n \, z^n \,, \qquad c_n \in V \,.
\eeq
In other words, $V \llbracket z, 1/z^2 \rrbracket$ can be identified
with the space of formal series $V \llbracket z, z^{-1} \rrbracket$.
The functional $\Res_z$ vanishes on a series \eqref{fs1d1}, while
the functional $\Res_z'$ coincides with the usual residue:
$\Res_z' \FF (z) = c_{-1}$.
\end{Example}

\subsection{Translation Invariance and Cauchy Formula}\label{se-cau}\bsec
%
Let $V$ be a vector space, as before.
One of the most important  properties of the residue map \eqref{eqn5}
is its translation invariance.

\skp

\begin{Proposition}\label{prp3n2} {\rm (Formal translation invariance.)}
\beq\label{eqn3n.5}
\Res_{\z} \ \iota_{\z,\w} \, \FF ( \z + \w ) \, = \,
\Res_{\z} \, \FF ( \z ) \,,
\qquad \FF ( \z ) \in V \llbracket \z \rrbracket_{(\z^2)^{\RR}}
\, . \
\eeq
This equation is also valid for elements of\/
$V \llbracket \z, (\z^2)^{\RR} \rrbracket$.
\end{Proposition}

\skp

\begin{Proof}
Recall that, by definition (cf.\ (\ref{eq2.21n2})),
$\iota_{\z,\w} \FF ( \z + \w ) = e^{\w \spr \Di_{\z}} \FF ( \z )$.
Then Eq.\ \eqref{eqn3n.5} follows from \ref{prp2n.2}(a).\end{Proof}

\skp

We proceed to finding 
an analog of the Cauchy kernel for our residue functional.

\skp

\begin{Proposition}\label{prp3n.1}
For\/
\(\psi ( \z ) \in V \llbracket \z \rrbracket \),
$\gamma \in \RR$, $n \in \NN$,
we have{\rm:}
\begin{align}\label{eqn3n.1}
\Res_{\z} \, ( \z^2 )^{\gamma} \, \psi ( \z ) &= 0
\qquad \text{if} \quad
\gamma \hspace{-1pt} + \hspace{-1pt} \frac{D}{2} > 0
\quad \text{or} \quad
\gamma \hspace{-1pt} + \hspace{-1pt} \frac{D}{2} \notin \ZZ
\, , \
\\ \label{eqn3n.2}
\Res_{\z} \, ( \z^2 )^{- \frac{D}{2}} \, \psi ( \z ) &=
\psi ( \zero ) \,,
\intertext{and}
\label{eqn3n.2-2}
\qquad
\Res_{\z} \, ( \z^2 )^{-\frac{D}{2}-n} \, \psi ( \z ) &=
K_n^{-1}
\bigl(( \Di_{\z}^2 )^n \psi \bigr) ( \zero ) \,,
\quad
\end{align}
where
\beq\label{kncau}
K_n := 2^{2n}
\mathop{\prod}\limits_{k = 1}^{n}
k
\left(
k \hspace{-1pt} + \hspace{-1pt} \frac{D}{2} \hspace{-1pt} - \hspace{-1pt} 1
\right) \,.
\eeq
\end{Proposition}

\skp

\begin{Proof}
Eqs.~(\ref{eqn3n.1}) and~(\ref{eqn3n.2})
are straightforward from the definition of the residue functional.
To prove (\ref{eqn3n.2-2}), it is enough to assume that
$\psi( \z )$ is a homogeneous polynomial of degree $2n$.
Then we apply induction on $n$,
starting with (\ref{eqn3n.2}), 
and using the $\Di_{\z}$--invariance \eqref{eqn6}
and the relation
\beqa
&& \nonumber
\mathop{\sum}\limits_{\al = 1}^{D} \,
\di_{z^{\al}}
\bigl( (\z^2 )^{-\frac{D}{2}-n+1}
( \di_{z^{\al}} \psi ) ( \z ) \bigr)
 \hspace{-10pt} \\ && \qquad\quad =
- 4 \, n \Bigl( 
n \hspace{-1pt} + \hspace{-1pt} \frac{D}{2} \hspace{-1pt} - \hspace{-1pt} 1
\Bigr) ( \z^2 )^{- \frac{D}{2}-n} \psi( \z )
+
( \z^2 )^{- \frac{D}{2}-n+1}
( \Di_{\z}^2 \psi) ( \z )
\,.
\nonumber \hspace{-10pt}
\eeqa
This completes the proof.\end{Proof}

\skp

As a corollary of \ref{prp3n.1},
for \textit{even} 
$D$
we have a local formula for the residue of an element
\(\FF (\z) \in V \llbracket \z \rrbracket_{\z^2} \):
\beq\label{loc_res}
\Res_{\z} \, \FF (\z) \, = \, K_N^{-1}
(\Di_{\z}^2)^N \bigl( (\z^2)^{N+\frac{D}{2}} \FF (\z)
\bigr) \big|_{\z=\zero} \,,
\qquad N \gg 0 \,.
\eeq

\skp

\begin{Proposition}\label{prp3nn3}
{\rm (Higher--dimensional ``Cauchy formula.'')}
\beq\label{eqn3nn6}
\Res_{\z} \, \iota_{\z,\w} \, 
\bigl( ( \z \hspace{-1pt} - \hspace{-1pt} \w )^2 \bigr)^{-\frac{D}{2}}
\, \psi(\z)
\, = \,
\psi(\w)
\qquad \text{for} \quad 
\psi(\z) \in V \llbracket \z \rrbracket
\, . \
\eeq
\end{Proposition}

\skp

\begin{Proof}
Consider the formal series
\[
\FF(\z,\w) := \iota_{\z,\w} \, 
\bigl( ( \z \hspace{-1pt} - \hspace{-1pt} \w )^2 \bigr)^{-D/2}
\, \psi(\z)
\in  V \llbracket \z \rrbracket_{\z^2} \,
\llbracket \w \rrbracket \,.
\]
By \eqref{eqn3n.5}, we have
\[
\Res_\z \, \FF(\z,\w) \, = \, \Res_\z \, \iota_{\z,\w'} \, \FF(\z+\w',\w) \,.
\]
We can put $\w'=\w$ in the right--hand side and obtain
\[
\Res_\z \, \iota_{\z,\w} \, \FF(\z+\w,\w)
\, = \, \Res_\z \, \iota_{\z,\w} \, (\z^2)^{-D/2} \, \psi(\z+\w)
\, = \, \psi(\w) \,,
\]
using \eqref{eqn3n.2}.\end{Proof}

\skp

\subsection{Harmonic Decomposition of\/ 
$\iota_{\z,\w} ((\z - \w)^2)^{\gamma}$}\label{se-hdg}\bsec
As before, let $\{ h_{m,\si}(\z) \}$
be an \emph{orthonormal} basis of the space 
of harmonic homogeneous polynomials of degree $m$
(see \eqref{eqn2n1}).
Introduce the polynomials
\beq\label{e3.13b}
\HR{m} (\z,\w) \, := \,
\mathop{\sum}\limits_{\sigma \, = \, 1}^{\har_m} \,
h_{m,\sigma} (\z) \, h_{m,\sigma} (\w) \,.
\eeq
Note that $\HR{m} (\z,\w)$ is the unique, up to a multiplicative constant,
$\Oo (D)$--invariant polynomial that is separately harmonic and  homogeneous
in $\z$ and $\w$ of degree $m$.
Combining Eq.~\eqref{eqn2n1} with the Cauchy formula \eqref{eqn3nn6},
we obtain that
\beq\label{e3.13a}
\iota_{\z,\w} \, 
\bigl( ( \z \hspace{-1pt} - \hspace{-1pt} \w )^2 \bigr)^{-\frac{D}{2}}
\, = \,
\mathop{\sum}\limits_{m,\, n \, = \, 0}^{\infty} \,
( \z^2 )^{-\frac{D}{2}-m-n} \, ( \w^2 )^n \, \HR{m} (\z,\w)
\,.
\eeq
Recall that for every $\ga\in\RR$ and $n\in\NNN$ the binomial
coefficient ${ \ga \choose n}$ is defined as 
$\ga(\ga-1) \dotsm (\ga-n+1) / n!$
and is a polynomial of $\ga$ of degree $n$.

\skp

\begin{Proposition}\label{pr-he}
For every $\ga\in\RR$, we have
\beq\label{e3.13}
\iota_{\z,\w} \,
\bigl( (\z \hspace{-1pt} - \hspace{-1pt} \w)^2 \bigr)^\ga
= \mathop{\sum}\limits_{m,\, n \, = \, 0}^{\infty} \,
K_{m,n}(\gamma) \,
( \z^2 )^{\gamma-m-n} \, ( \w^2 )^n \, \HR{m} (\z,\w)
\,,
\eeq
where 
\beq\label{kmng}
K_{m,n}(\gamma) \, := \, 
\frac{(-1)^n}{ { -\frac{D}{2} \choose m+n} }
\, { \frac{D}{2} -1+\ga \choose n}
\, { \ga \choose m+n} \,.
\eeq
\end{Proposition}

\skp

\begin{Proof}
It follows from definition \eqref{eq2.21n2} and the $\Oo (D)$--invariance
that $\iota_{\z,\w} \hspace{1pt} ((\z - \w)^2)^{\gamma}$
has the form \eqref{e3.13}. Moreover, it is clear from \eqref{eq2.21n2}
that for each fixed $m,n \in \NNN$, the coefficient $K_{m,n}(\gamma)$
is a polynomial of $\gamma$. Then to establish \eqref{kmng}
it will suffice to prove it for infinitely many values of $\ga$.

We will prove by induction that formula \eqref{kmng} holds
for all $\ga$ such that $-\frac{D}{2} - \ga \in \NNN$.
For $\ga = -\frac{D}{2}$ it gives $K_{m,n}(-\frac{D}{2}) = 1$,
which agrees with \eqref{e3.13a}. Next, assume that
\eqref{e3.13} and \eqref{kmng} hold for some $\ga$.
Apply the Laplace operator $\Di_\z^2$ to both sides of
\eqref{e3.13} and use \eqref{ddf2} to find:
\begin{align*}
\ga & \bigl( \ga+ \tfrac{D}{2} - 1 \bigr) \,
\iota_{\z,\w} \,
\bigl( (\z \hspace{-1pt} - \hspace{-1pt} \w)^2 \bigr)^{\ga-1}
\\
&= \mathop{\sum}\limits_{m,\, n \, = \, 0}^{\infty} \,
K_{m,n}(\gamma) \,
(\gamma-m-n) \bigl( \ga -n + \tfrac{D}{2} - 1 \bigr) \,
( \z^2 )^{\gamma-1-m-n} \, ( \w^2 )^n \, \HR{m} (\z,\w)
\,.
\end{align*}
Comparing this to \eqref{e3.13} with $\ga-1$ instead of $\ga$,
we obtain that \eqref{kmng} holds for $\ga-1$.
This completes the proof.\end{Proof}

\skp

For $\ga=-\frac{D}{2}+1$, expansion \eqref{e3.13}
takes the particularly simple form
\beq\label{e3.13d}
\iota_{\z,\w} \, 
\bigl( (\z-\w)^2 \bigr)^{-\frac{D}{2}+1}
\, = \,
\mathop{\displaystyle\sum}\limits_{m \, = \, 0}^{\infty} \,
\frac{\frac{D}{2}-1}{\frac{D}{2}-1+m} \,
(\z^2)^{-\frac{D}{2}+1-m} \, \HR{m} (\z,\w) \,.
\eeq
Note that both sides of this equation
are harmonic with respect to both $\z$ and $\w$ (see \ref{rm-harm}).
Let us also point out that for fixed $\ga\in\NNN$, the coefficient
$K_{m,n}(\gamma)$ vanishes whenever $m+n > \ga$. 
Using \eqref{e3.13} for $\ga=0,1,2,\dots$, one can find the polynomials
$\HR{m} (\z,\w)$ recursively; for example,
\beq\label{eq-hrm}
\begin{split}
\HR{0} (\z,\w) \, = \, 1 & \,, \qquad
\HR{1} (\z,\w) \, = \, D \, \z\spr\w \,,
\\
\HR{2} (\z,\w) \, &= \, \bigl( \tfrac{D}{2} + 1 \bigr) 
\bigl( D (\z\spr\w)^2 - \z^2 \, \w^2 \bigr) \,.
\end{split}
\eeq

\subsection{Formal Delta--Function}\label{se-de}\bsec
In this subsection we define a formal distribution in two variables,
which plays the role of the delta--distribution.
Let us consider the $\ZZ$--invariant set 
$\ZZ' := \frac{D}{2} + \ZZ$
(cf.\ \eqref{eq2.36n3}), which coincides with $\ZZ$ when $D$ is even.
We define the following formal distribution in two variables
\beq\label{eq-de1}
\de(\z,\w) := \,  
\mathop{\sum}\limits_{n \in \ZZ} \,
\mathop{\sum}\limits_{m \, = \, 0}^{\infty} \,
( \z^2 )^{-\frac{D}{2}-m-n} \, ( \w^2 )^n \, \HR{m} (\z,\w)
\; \in \CC\llbracket \z, (\z^2)^{\ZZ'} ; \w, (\w^2)^\ZZ \rrbracket
\,.
\eeq


\skp

\begin{Proposition}\label{pr-de}
The above-defined\/ $\de(\z,\w)$ is the unique element of\/
$\CC\llbracket \z,$ $(\z^2)^{\ZZ'};$ $\w,$ $(\w^2)^\ZZ \rrbracket$
with the property that
\beq\label{eq-de3}
\Res_\z \, \FF(\z) \, \de(\z,\w) = \FF(\w)
\qquad \text{for all} \quad \FF(\z) \in \CC[\z]_{\z^2} \,.
\eeq
In addition, it satisfies\/
\beq\label{eq-de4}
\FF(\z) \, \de(\z,\w) = \FF(\w) \, \de(\z,\w) \,,
\qquad \FF(\z) \in \CC[\z]_{\z^2}
\eeq
and
\beq\label{eq-de5}
\di_{z^\al} \, \de(\z,\w) = -\di_{w^\al} \, \de(\z,\w) \,,
\qquad \al=1,\dots,D \,.
\eeq
\end{Proposition}

\skp

\begin{Proof}
Property \eqref{eq-de3} and the uniqueness of $\de(\z,\w)$ follow
from the orthogonality relation \eqref{eqn2n1} (cf.\ \ref{pr2.3n3}).
Then Eq.~\eqref{eq-de3} and
the $\Di_{\z}$--invariance of the residue \eqref{eqn6} 
imply that, for every $\FF(\z) \in \CC[\z]_{\z^2}$,
\begin{align*}
\Res_\z \, \FF(\z) \, & (\di_{z^\al} + \di_{w^\al}) \, \de(\z,\w)
\\
& = \, -\Res_\z \, \bigl( \di_{z^\al} \FF(\z) \bigr) \, \de(\z,\w)
\, + \, \di_{w^\al} \Res_\z \, \FF(\z) \, \de(\z,\w) 
\, = \, 0 \,.
\end{align*}
This proves \eqref{eq-de5}.
Similarly, \eqref{eq-de4} follows from the equalities
\[
\Res_\z \, \psi(\z) \, \FF(\z) \, \de(\z,\w) =
\psi(\w) \, \FF(\w) =
\Res_\z \, \psi(\z) \, \FF(\w) \, \de(\z,\w) \,,
\]
for all $\psi(\z), \FF(\z) \in \CC[\z]_{\z^2}$.\end{Proof}

\skp

\begin{Remark}\label{rm-de1}
Let $\Gamma$ be any $\ZZ$--invariant subset of $\RR$,
and let $\Gamma' = - \Gamma + \frac{D}{2}$ (see \eqref{eq2.36n3}).
\ref{pr2.3n3} implies that there exists a unique element
$\de_\Ga(\z,\w) \in \CC\llbracket \z,$ 
$(\z^2)^{\Ga'};$ $\w,$ $(\w^2)^\Ga \rrbracket$
such that Eq.~\eqref{eq-de3} holds for $\de_\Ga(\z,\w)$
and all $\FF(\z) \in \CC[\z]_{(\z^2)^\Ga}$.
Then \eqref{eq-de5} is satisfied as well, while the analog
of \eqref{eq-de4} holds only when $\Ga$ is a subgroup of $\RR$
(this is needed for $\CC[\z]_{(\z^2)^\Ga}$ to be a ring).
Finally, note that $\de_\Ga(\z,\w) = \de_\Ga(\w,\z)$
when $\Ga=\Ga'$.
\end{Remark}

\skp

Observe that $\de(\z,\w)$ is symmetric,
i.e., $\de(\z,\w) = \de(\w,\z)$, if and only if $D$ is even.
Switching $\z$ and $\w$ in \eqref{e3.13a} and using
\eqref{zwgwzg}, we obtain
that for even $D$ we have:
\beq\label{eq-de2}
\begin{split}
\de(\z,\w) 
\, = & \,\,
\iota_{\z,\w} \, 
\bigl( ( \z \hspace{-1pt} - \hspace{-1pt} \w )^2 \bigr)^{-\frac{D}{2}}
+ \iota_{\w,\z} \, 
\bigl( ( \z \hspace{-1pt} - \hspace{-1pt} \w )^2 \bigr)^{-\frac{D}{2}}
\\
&+ 
\mathop{\sum}\limits_{m \, = \, 0}^{\infty} \,
\mathop{\sum}\limits_{n \, = \, 1}^{\frac{D}{2}-1+m} \,
( \z^2 )^{-\frac{D}{2}-m+n} \, ( \w^2 )^{-n} \, \HR{m} (\z,\w)
\,.
\end{split}
\eeq

This splitting of $\de(\z,\w)$ as a sum of three terms suggests the
introduction of a natural partition of the space of formal distributions
(without assuming that $D$ is even).
For a formal distribution $\FF(\z) \in V \llbracket z, 1/z^2 \rrbracket$,
written as in \eqref{fd10}, we define its parts
$\FF(\z)_+$,
$\FF(\z)_-$,
and 
$\FF(\z)_\sim$,
as follows.
We let $\FF(\z)_+$ be given by \eqref{fd10} with the sum over
$n \in \ZZ$ restricted to $n \geqslant 0$, $n \in \ZZ$.
For $\FF(\z)_-$ we restrict the sum to  
$n \leqslant -\frac{D}2 -m$, $n \in \ZZ$
(we first sum over $m$ and then over $n$).
For $\FF(\z)_\sim$ we restrict the sum to
$-\frac{D}2+1-m \leqslant n \leqslant -1$, $n \in \ZZ$.
%
%
We call $\FF(\z)_+$ the \textbf{regular part} of $\FF(\z)$,
and we define the \textbf{singular part} as
\beq\label{fd17}
\FF(\z)_\sip \, := \, \FF(\z)_- + \FF(\z)_\sim \,.
\eeq
Then, obviously, 
$\FF(\z)_+ \in V \llbracket \z \rrbracket$ and
\beq\label{fd18}
\FF(\z) \, = \,
\FF(\z)_+ + \FF(\z)_- + \FF(\z)_\sim \, = \,
\FF(\z)_+ + \FF(\z)_\sip \,.
\eeq
It is important that
the above partition of the space of formal distributions is 
$\CC[\Di_\z]$--invariant, i.e.,
\beq\label{fd22}
\bigl( \di_{z^\al} \FF(\z) \bigr)_\star
\, = \, \di_{z^\al} \, \FF(\z)_\star \,, \quad
\text{for} \quad \star=+,-,\sim,\sip
\quad \text{and} \quad \al=1,\dots,D \,. \;
\eeq

Let us point out that the product 
$\FF(\z) \, \iota_{\z,\w} \, ((\z-\w)^2)^{-\frac{D}{2}}$
is well defined and belongs to the space
$V \llbracket \z, 1/\z^2 \rrbracket \llbracket \w \rrbracket$.
Then
we can generalize Cauchy formula \eqref{eqn3nn6},
using Eq. \eqref{e3.13a}, to obtain
\beq\label{fd19}
\Res_{\w} \, \FF (\z) \ \iota_{\z,\w} \, 
\bigl((\z-\w)^2\bigr)^{-\frac{D}{2}} \, = \,
\FF (\w)_+  \,, \qquad
\FF(\z) \in V \llbracket \z, 1/\z^2 \rrbracket \,.
\eeq
Thus, $\iota_{\z,\w} \, \bigl((\z-\w)^2\bigr)^{-\frac{D}{2}}$ may be called
$\delta^+ (\z,\w)$, and one can also introduce formal distributions
$\delta^- (\z,\w)$ and $\delta^\sim (\z,\w)$ that give 
the $-$ and $\sim$ parts of $\FF(\w)$, respectively, and such that
$\delta (\z,\w) = \delta^+ (\z,\w) + \delta^- (\z,\w) + 
\delta^\sim (\z,\w)$.
In the case when $D$ is even, this splitting of $\de(\z,\w)$
coincides with the one in Eq.~\eqref{eq-de2}.

\skp

\begin{Remark}\label{rm-de2}
Introduce the formal distribution (cf.\ \eqref{e3.13a})
\[
\de_\harm^+(\z,\w) 
\, := \,
(\z^2 - \w^2) \,\,
\iota_{\z,\w}
\bigl( ( \z \hspace{-1pt} - \hspace{-1pt} \w )^2 \bigr)^{-\frac{D}{2}}
\, = \,
\mathop{\displaystyle\sum}\limits_{m \, = \, 0}^{\infty} \,
(\z^2)^{-\frac{D}{2}+1-m} \, \HR{m} (\z,\w) \,.
\]
It is harmonic with respect to both $\z$ and $\w$,
and has the property that
\[
\Res_\z \, h(\z) \, \de_\harm^+(\z,\w) = h(\w)
\qquad \text{for all} \quad h(\z) \in \CC[\z]^\harm \,.
\]
It follows from \eqref{eq-de1} that
\(
\de(\z,\w) \, = \, \de_1(\z^2-\w^2) \, \de_\harm^+(\z,\w),
\)
where
\[
\de_1(x-y) \, := \, 
\mathop{\sum}\limits_{n \in \ZZ} \, x^{-1-n} \, y^n 
\in \CC \llbracket x,x^{-1} ; y,y^{-1} \rrbracket
\]
is the formal delta--distribution in the usual $D=1$ theory of 
vertex algebras (see e.g.\ \cite{FLM,K,LL}). Notice that, even though
$(\z^2 - \w^2) \, \de_1(\z^2-\w^2) = 0$, one can \emph{not} conclude
from here that $\de(\z,\w) = 0$, because the product
$\de_1(\z^2-\w^2) \,\, \iota_{\z,\w} \bigl( (\z-\w)^2 \bigr)^{-\frac{D}{2}}$
is not well defined (see the discussion in \cite[Sect.\ 2.1]{LL}
and in particular Eq.~(2.1.17)).
\end{Remark}

\subsection{Transformation Properties}\label{se-res-tr}\bsec
For completeness,
in this subsection we will investigate the transformation properties of
the residue functional and the $\iota$--operation under
the \textit{conformal inversion}
$\z \mapsto \z / \z^2$.

We observe that the substitution
$\FF(\z) \mapsto \FF( \z / \z^2 )$
is a well-defined involution of
$V \llbracket \z, (\z^2)^{\RR} \rrbracket$ if we set
$\bigl(( \z / \z^2 )^2 \bigr)^\ga := ( \z^2 )^{-\ga}$.
Explicitly, if $\FF(\z)$ is given by \eqref{fd11}, then
\begin{equation}\label{fd13}
\FF \Bigl( \frac{\z}{\z^2} \Bigr)
\, := \, 
\sum_{\ga\in\Ga} \, \sum_{m=0}^\infty \, \sum_{\si=1}^{\har_m} \,
\FF_{\{\ga,m,\si\}} \, (\z^2)^{-\ga-m} \, h_{m,\si}(\z) 
\,, 
\end{equation}
because $h_{m,\si}( \z / \z^2 ) = (\z^2)^{-m} \, h_{m,\si}(\z)$.
Clearly, under this isomorphism, the $\CC [\z]_{\z^2}$--module
$V\llbracket \z,(\z^2)^{\Gamma} \rrbracket$ is mapped onto 
$V\llbracket \z,(\z^2)^{-\Gamma} \rrbracket$.

Now let $\Gamma$ be
an
additive subgroup of $\RR$ 
containing $\ZZ$. For $\ga\in\Ga$, we define
\beq\label{eqn3n8}
\Bigl(\Bigl(
\frac{\z}{\z^2} - \frac{\w}{\w^2}
 \Bigr)^2 \, \Bigr)^\ga
\, := \,
\bigl( \z^2 \bigr)^{-\gamma} \bigl( \w^2 \bigr)^{-\gamma}
\bigl( (\z-\w)^2 \bigr)^{\gamma}
\,,
\eeq
which agrees with the usual formula for $\ga=1$.
Then the substitution
$\psi(\z,\w) \mapsto \psi( \z / \z^2 , \w / \w^2 )$
defines an automorphism of
$V [ \z,\w ]_{(\z^2)^\Gamma \com (\w^2)^\Gamma
\com ((\z-\w)^2)^\Gamma}$.


\skp

\begin{Proposition}\label{prp3n3}
Let\/ 
\(\FF ( \z ) \in V \llbracket \z,(\z^2)^{\RR} \rrbracket\) 
and\/
\( \psi( \z,\w ) \in
V [ \z,\w ]_{(\z^2)^\Gamma \com (\w^2)^\Gamma
\com ((\z-\w)^2)^\Gamma} \).
Then we have{\rm:}
\beq\label{eqn3n9}
\Res_{\z} \, \FF \left(\raisebox{10pt}{\hspace{-3pt}}\right.
\frac{\z}{\z^2}
\left.\raisebox{10pt}{\hspace{-3pt}}\right)
\, = \,
\Res_{\z} \, (\z^2)^{-D} \, \FF ( \z )
\eeq
and
\beq\label{eqn3n10}
\iota_{\z,\w}
\left(\raisebox{10pt}{\hspace{-3pt}}\right.
\psi \left(\raisebox{10pt}{\hspace{-3pt}}\right.
\frac{\z}{\z^2},\frac{\w}{\w^2}
\left.\raisebox{10pt}{\hspace{-3pt}}\right)
\left.\raisebox{10pt}{\hspace{-3pt}}\right)
\, = \,
\left(\raisebox{10pt}{\hspace{-3pt}}\right.
\iota_{\w',\z'} \, \psi( \z',\w' )
\left.\raisebox{10pt}{\hspace{-3pt}}\right)
\Big|_{ \z' = \z / \z^2 , \, \w' = \w / \w^2  }
\, . \
\eeq
\end{Proposition}

\skp

\begin{Proof}
Eq.~(\ref{eqn3n9}) is immediate from the definition of the residue
(cf.\ \eqref{fd11}, \eqref{fd13}).
To prove (\ref{eqn3n10}), it suffices to check it for 
\( \psi ( \z,\w ) = \bigl( (\z - \w)^{2} \bigr)^{\gamma} \),
in which case the statement follows from \eqref{zwgwzg}, \eqref{e3.13}
and~\eqref{eqn3n8}.\end{Proof}

\skp

One can define involutive automorphisms 
\beq\label{eq-z1}
\FF(\z) \, \mapsto \,  \FF( -\z ) \,, \qquad
\psi(\z,\w) \, \mapsto \, \psi( -\z, -\w )
\eeq
of
$V \llbracket \z, (\z^2)^{\RR} \rrbracket$
and $V \llbracket \z, \w \rrbracket_{(\z^2)^\Gamma \com (\w^2)^\Gamma
\com ((\z-\w)^2)^\Gamma}$ by setting (cf.\ \eqref{zwgwzg})
\beq\label{eq-z2}
\bigl( \z^2 \bigr)^{\ga_1} \bigl( \w^2 \bigr)^{\ga_2} 
\bigl( (\z-\w)^2 \bigr)^{\ga_3} f(\z,\w)
\, \longmapsto \, 
\bigl( \z^2 \bigr)^{\ga_1} \bigl( \w^2 \bigr)^{\ga_2} 
\bigl( (\z-\w)^2 \bigr)^{\ga_3} f(-\z,-\w)
\eeq
for $f(\z,\w) \in V \llbracket \z, \w \rrbracket$.
By the definition of $\iota_{\z,\w}$ and $\iota_{\w,\z}$
(see Sect.~\ref{se1.2}), the so-defined operation commutes with both
of them. It is also clear that it anti-commutes with all partial
derivatives $\di_{z^\al}$, $\di_{w^\al}$, and satisfies
\beq\label{eq-z3}
\Res_\z \, \FF( -\z ) \, = \, \Res_\z \, \FF( \z ) \,, \qquad
\FF(\z) \in V \llbracket \z, (\z^2)^{\RR} \rrbracket \,.
\eeq

\skp

\begin{Remark}\label{rm-z-z}
In analogy with
the above automorphism \eqref{eq-z2},
one can
define an automorphism $\Theta$ by setting 
\begin{align*}
\Theta \colon 
\bigl( \z^2 \bigr)^{\ga_1} \bigl( \w^2 \bigr)^{\ga_2} &
\bigl( (\z-\w)^2 \bigr)^{\ga_3} f(\z,\w)
\\
& \longmapsto \, 
e^{ 2\pi i (\ga_1+\ga_2+\ga_3) } 
\bigl( \z^2 \bigr)^{\ga_1} \bigl( \w^2 \bigr)^{\ga_2} 
\bigl( (\z-\w)^2 \bigr)^{\ga_3} f(\z,\w)
\end{align*}
for $f(\z,\w) \in V \llbracket \z, \w \rrbracket$.
Then $\Theta$ commutes with the $\iota$--operations and
with the partial derivatives, 
and instead of \eqref{eq-z3} one has{\rm:}
$\Res_\z \, (\Theta\FF)( \z )$ $=$ $(-1)^D \Res_\z \, \FF( \z )$.
\end{Remark}

\skp

\section{Fields and Locality}\label{se3}
\setcntrs

In this section we investigate the notions of fields, locality
and operator product expansion in higher dimensions,
mainly following \cite{N05}.
We give the definition of vertex algebra and we provide two examples
of vertex algebras. 

\subsection{Polylocal Fields}\label{se3.1}\bsec
In this subsection we introduce the notion of a field of several variables,
and we generalize the results of \cite[Sect.~2]{N05}
about the existence of operator product expansion.

Let $V = V_{\bar 0} \oplus V_{\bar 1}$ be a $\ZZ_2$--graded vector space
(i.e., a \emph{superspace}).
Then $\End V$ $=$ $(\End V)_{\bar 0}$ $\oplus$ $(\End V)_{\bar 1}$
is a $\ZZ_2$--graded associative algebra, and we will denote
its Lie super bracket by
\beq\label{eq3.1n2}
[A,B] \, := \, AB - (-1)^{pq} BA \,, \qquad\text{for} \;\;
A \in (\End V)_p \,, \;\; B \in (\End V)_q \,.
\eeq
We will suppose that $V$ is endowed with an action
of mutually commuting even endomorphisms $T_1,\dots,T_D$
(called \textbf{translation endomorphisms})
and with an even vector $\vac$ (called \textbf{vacuum}), such that
$T_1 \vac = \dots = T_D \vac = 0$.

Let $A (\z_1,\dots,\z_m)$ be an $(\End V)$--valued formal distribution;
in other words, let
\beq
A (\z_1, \dots, \z_m) \in
(\End V) \llbracket \z_1, 1/\z_1^2; \dots;
\z_m, 1/\z_m^2 \rrbracket \,.
\eeq
It is called a \textbf{field} in $\z_1,\dots,\z_m$ (or just an
$m$--\textbf{field}) iff for every $v \in V$ one has
\beq\label{eq3.2n2}
A (\z_1,\dots,\z_m) \, v \, \in \,
V \llbracket \z_1,\dots,\z_m \rrbracket_{\z_1^2\dotsm\z_m^2}.
\eeq
This means that for every $v \in V$ there is 
a non-negative integer $N_{A,v}$ such that
\beq\label{eq3.4n2}
A (\z_1,\dots,\z_m) \, v \, = \,
(\z_1^2 \dotsm \z_m^2)^{-N_{A,v}} \,
\psi_{A,v} (\z_1,\dots,\z_m)
\eeq
for some \(\psi_{A,v} (\z_1,\dots,\z_m)
\in V \llbracket \z_1,\dots,\z_m \rrbracket\).

If $A$ is an $m$--field, then for every partition
\beq\label{eq3.5n2}
\{1,\dots,m\} \, = \,
J_1 \sqcup \dotsm \sqcup J_r
\qquad \text{(disjoint union)} , 
\eeq
the \textbf{restriction}
\beq\label{eq3.6n2}
\widetilde A (\u_1,\dots,\u_r) \, v \, := \,
\Bigl( 
(\z_1^2 \dotsm \z_m^2)^{-N_{A,v}} \,
\psi_{A,v} (\z_1,\dots,\z_m)
\Bigr) \Big|_{ \, \z_j := u_s \,\;\text{for}\,\; j \in J_s }
\eeq
makes sense and defines again a field.


An $m$--field (or, more generally, an $(\End V)$--valued formal distribution)
$A$ is called \textbf{translation invariant} iff
\beq\label{eq3.7n2}
[T_{\al},A (\z_1,\dots,\z_m)] \, = \,
\mathop{\sum}\limits_{k \, = \, 1}^m \,
\di_{z^{\al}_k} \, A (\z_1,\dots,\z_m)
\eeq
for every $\al = 1,\dots,D$.

Let us point out that a product
$A (\z_1,\dots,\z_m) B (\w_1,\dots,\w_n)$
of two fields 
is \emph{not} a field in general.
Indeed, by the above definition, for every $v \in V$ we have
\beq\label{eq3.8n3}
A (\z_1\hspace{-1pt},\hspace{-1pt}\dots\hspace{-1pt},\hspace{-1pt}\z_m)
\hspace{2pt}
B (\w_1\hspace{-1pt},\hspace{-1pt}\dots\hspace{-1pt},\hspace{-1pt}\w_n)
\hspace{2pt} v
\hspace{0pt} \in \hspace{0pt}
V \llbracket \z_1\hspace{-1pt},\hspace{-1pt}\dots\hspace{-1pt},
\hspace{-1pt}\z_m
\rrbracket_{\z_1^2\dotsm\z_m^2}
\,
\llbracket \w_1\hspace{-1pt},\hspace{-1pt}\dots\hspace{-1pt},\hspace{-1pt}\w_n
\rrbracket_{\w_1^2\dotsm\w_m^2}
\,,
\eeq
and it may contain infinitely many negative powers of 
$\z_1^2,\dots,\z_m^2$.
As a consequence, the restriction of the product \eqref{eq3.8n3}
for coinciding arguments is not well defined in general.
We will show below that one can ``regularize'' this product
to make a field if the following definition is satisfied.

Two fields
(or, more generally, $(\End V)$--valued formal distributions)
$A$ and $B$
are called
\textbf{mutually local} iff there exists a non-negative
integer $N_{A,B}$ such that
\beq\label{eq3.8n2}
\Bigl(
\mathop{\prod}\limits_{j \, = \, 1}^m
\mathop{\prod}\limits_{k \, = \, 1}^n
(\z_j-\w_k)^2
\Bigr)^{\! N_{A,B}} \,
\bigl[ A (\z_1,\dots,\z_m),B (\w_1,\dots,\w_n) \bigr]
\, = \, 0 \,.
\eeq
A $1$--field that is local with respect to itself is usually called
a \textbf{local} field;
a $2$--field that is local with respect to itself is called
a \textbf{bilocal} field.
An $m$--field, for general $m$, which is local with respect to itself,
is called a \textbf{polylocal} field.

In the following statement we sum up some consequences
of the above definitions.

\skp

\begin{Theorem}\label{pr3.1n2}
Let $A (\z_1,\dots,\z_m)$ and $B (\z_1,\dots,\z_n)$ be an $m$--field
and an $n$--field, respectively, which are mutually local as above.

\medskip

{\rm (a)}
Every restriction \eqref{eq3.6n2} of
$A$ 
is also a field and
is mutually local with respect to~$B$. 

\medskip

{\rm (b)}
If the field $A$ 
is translation invariant, then
its restrictions are also translation invariant fields.

\medskip

{\rm (c)}
If $A$ 
is translation invariant, then
$A (\z_1, \dots, \z_m) \vac \in
V \llbracket \z_1, \dots, \z_m \rrbracket$.

\medskip

{\rm (d)}
Every partial derivative $\di_{z^{\al}_k} A$
is a field and is mutually local with respect to $B$. 
If the field $A$ is translation invariant, then
$\di_{z^{\al}_k} \, A$ is also translation invariant.

\medskip

{\rm (e)}
The formal distribution
\beq\label{eq3.9n2}
\begin{split}
F_{A,B} &(\z_1,\dots,\z_m,\w_1,\dots,\w_n) \,
\\
& := \,
\Bigl(
\mathop{\prod}\limits_{j \, = \, 1}^m
\mathop{\prod}\limits_{k \, = \, 1}^n
(\z_j-\w_k)^2
\Bigr)^{\! N_{A,B}} \,
A (\z_1,\dots,\z_m) \, B (\w_1,\dots,\w_n)
\end{split}
\eeq
is an $(m+n)$--field. If the fields $A$ 
and $B$ 
are local with respect to a $p$--field $C (\z_1,$ $\dots,$ $\z_p)$,
then $F_{A,B}$ is also local with respect to $C$.
If both fields $A$ and $B$ are translation invariant, then
$F_{A,B}$ is also translation invariant.
\end{Theorem}

\skp

\begin{Proof}
Statements
(a) and (b) follow easily from definitions.

Statement (c) for $m = 1$ is proved in \cite[Proposition~3.2(a)]{NT05},
and that proof can be straightforwardly generalized for general $m$
(note that one can take $h_2=h_1$ there).

To prove (d), one ``commutes'' the derivative $\di_{z^{\al}_k}$
through the polynomial $((\z_k-\w_j)^2)^N$, as it is done in a more general
case in \cite[Lemma 2.3]{N05}.

(e) Note that, by (\ref{eq3.8n3}) and (\ref{eq3.8n2}),
for every $v \in V$ the series
$F_{A,B}(\z_1, \dots, \z_m$, $\w_1, \dots, \w_n) \, v$
belongs to the intersection
\[
V \llbracket \z_1, \dots, \z_m \rrbracket_{\z_1^2\dotsm\z_m^2}
\llbracket \w_1, \dots, \w_n \rrbracket_{\w_1^2\dotsm\w_n^2}
\, \cap \,
V \llbracket \w_1, \dots, \w_n \rrbracket_{\w_1^2\dotsm\w_n^2}
\llbracket \z_1, \dots, \z_m \rrbracket_{\z_1^2\dotsm\z_m^2}
\]
which is exactly
$V \llbracket \z_1, \dots, \z_m, \w_1, \dots,
\w_n \rrbracket_{\z_1^2\dotsm\z_m^2\w_1^2\dotsm\w_n^2}$.
But this means, by definition, that $F_{A,B}$ is an $(m+n)$--field.
The remaining part of the statement is straightforward.\end{Proof}

\skp

As a corollary of \ref{pr3.1n2},
every $m$--field $A (\z_1,\dots,\z_m)$ can be expanded
in $1$--fields as follows.
Consider for $v \in V$ the formal expansion
\begin{align}
\notag
\iota_{\z,\w_1} & \dotsm \, \iota_{\z,\w_{m-1}} \,
A (\z+\w_1,\dots,\z+\w_{m-1},\z) \, v
\\
\label{eq3.10n2}
& := \,
\exp( \w_1 \spr \Di_{\z_1} + \dotsm + \w_{m-1} \spr \Di_{\z_{m-1}} )
\,\, A (\z_1,\dots,\z_m) \, v \,
\big|_{ \z_1 = \dotsm = \z_m = \z  }
\\ 
\notag
& \in \,
V \llbracket \z \rrbracket_{\z^2}
\,
\llbracket \w_1,\dots,\w_{m-1} \rrbracket \,.
\end{align}
This is a formal power series in $\w_1,\dots,\w_{m-1}$ with
coefficients of the form
\(\psi_i (\z) \, v \in V \llbracket \z \rrbracket_{\z^2}\)
for some uniquely defined fields $\psi_i (\z)$
($i$ running over some index set).
All $\psi_i (\z)$ are fields because
they are obtained from $A (\z_1,\dots,\z_m)$ by the operations
of differentiation and restriction.
If, in addition,
$A$ is translation invariant and is local with respect to
some other fields $B$, $C$, etc., then all the fields $\psi_i (\z)$
are also translation invariant and local with respect to $B$, $C$, etc.

The formal expansion~(\ref{eq3.10n2}) is called
the \textbf{operator expansion} of $A (\z_1,\dots,\z_m)$.
Applying this expansion to the field $F_{A,B}$~(\ref{eq3.9n2}),
we get what is called the \textbf{operator product expansion} (OPE) of two
mutually local fields $A$ and~$B$.

\skp

\begin{Example}\label{rm3.1n4}
Let us consider, for comparison, the $D=1$ case of OPE.
Recall from \ref{exresd1} that now 
$\z \! = \! z$ is a $1$-component variable
and the space of $(\End V)$--valued formal distributions is identified with
$(\End V) \llbracket z, z^{-1} \rrbracket$.
Then our notions of \emph{fields} and \emph{locality} coincide with the
ones used in vertex algebra theory 
(see \cite{G,DL,L1,K,LL}).
For two mutually local fields $a (z)$ and $b (z)$
with parities $p_a$ and $p_b$, respectively, one introduces their
\textbf{$n$-th product} for $n \in \ZZ$ by
\beq\label{eq3.11n3}
\begin{split}
\bigl( a(w)_{(n)}b(w) \bigr) c \, := \,
& \RES_z \, a (z) \, b (w) \, c \ \iota_{z,w} \, (z-w)^n
\\
&- (-1)^{p_a \, p_b}
\RES_z \, b (w) \, a (z) \, c \ \iota_{w,z} \, (z-w)^n \,,
\end{split}
\eeq
where $\RES_z \, z^k := \delta_{k,-1}$ is the usual residue functional
(it corresponds to our $\Res_z'$; see \ref{exresd1}).
By the Cauchy theorem for $\RES_z$ (see \cite{FLM,K,LL}),
one gets an equivalent definition
\beq\label{eq3.12n3}
\bigl( a(w)_{(n)}b(w) \bigr) c \, = \,
\frac{1}{N!} \, \di_{z}^N \Bigl(
(z-w)^{N+n+1} \, a (z) \, b (w) \, c \Bigr) \Big|_{z = w} \,,
\quad N\gg0 \,, \;
\eeq
where the right--hand side is independent of $N$.
Our approach to the OPE of $a (z)$ and $b (z)$
corresponds precisely to definition \eqref{eq3.12n3}.
Then \emph{Dong's Lemma}, the fact that the field
$a(z)_{(n)}b(z)$ is local with respect
to every field $c(z)$ local with respect to $a$ and $b$, is a simple
corollary of \ref{pr3.1n2} (e), (d), (a).
\end{Example}

\subsection{Completeness and State--Field Correspondence.
        Definition of Vertex Algebra}\label{se3.2}\bsec
The translation invariance and locality properties allow us
to introduce a state--field correspondence
for a vertex algebra in higher dimensions, 
as in Sect.\ 3 and 4 of \cite{N05}.
Here we will reproduce these results in a more concise way.

As in the previous subsection, 
let $V$ be a superspace endowed with an action of
mutually commuting even endomorphisms $T_{\al}$ ($\al = 1,\dots,D$)
and a vacuum vector~$\vac$.
A system of fields $\{ \phi_{i} (\z) \}_{ i \in \mathcal{I} }$
is called \textbf{local} iff
$\phi_{i} (\z)$ and $\phi_{j} (\z)$ are mutually local
for every $i,j \in \mathcal{I}$.
The system $\{\phi_{i} (\z)\}$ is called
\textbf{translation invariant}
iff every $\phi_{i} (\z)$ is translation invariant.
Finally, the system $\{\phi_{i} (\z) \}$
is called \textbf{complete} (with respect to the vacuum $\vac$) iff
the coefficients of all formal series
\(\phi_{i_1} (\z_1) \dotsm \phi_{i_n} (\z_n) \, \vac\)
($n\in\NN$) together with $\vac$ span the whole vector space $V$.
In other words, the system $\{\phi_{i} (\z) \}$
is complete iff the vacuum $\vac$ is a \textit{cyclic vector} for the
associative subalgebra of $\End V$
generated by the modes of all fields $\phi_{i} (\z)$.

\skp

\begin{Theorem}\label{th3.2n2}
Let\/ $\{ \phi_{i} (\z) \}_{ i \in \mathcal{I} }$ be a
translation invariant, local and complete system of fields.
Then for every $a \in V$ there exists a unique field, denoted as
$Y (a,\z)$, which is translation invariant, local with respect to
all $\phi_{i} (\z)$, and such that
\beq\label{eq3.11n2}
Y (a,\z) \, \vac \big|_{\z = \zero} \, = \, a \,.
\eeq
\end{Theorem}

\skp

\begin{Proof}
Let us consider the vector space $\mathcal{F}$ of all translation invariant
$1$--fields that are local with respect to $\phi_{i} (\z)$
for all $i \in \mathcal{I}$.
By \ref{pr3.1n2}(c), there is a well-defined linear map
\beq\label{eq3.12n2}
\mathcal{F} \to  V \,, \qquad
\chi (\z) \mapsto \chi (\z) \, \vac \, \big|_{\z = \zero} \,.
\eeq
It follows from translation invariance that
\[
\chi (\z) \, \vac \, = \, e^{\z \spr \T}
\bigl(
\chi (\w) \, \vac \big|_{\w = \zero}
\bigr) \,,
\qquad\quad
\z \spr \T := z^1 T_1 + \dots + z^D T_D \,.
\]
Then Theorem~3.1 from \cite{N05} implies that map 
(\ref{eq3.12n2}) is injective.
The theorem will be proved as soon as we
show that map (\ref{eq3.12n2}) is surjective.

Consider for every fixed
$m = 1,2,\dots$ and $i_1,\dots,i_m \in \mathcal{I}$
the $m$--field
\[
A (\z_1,\dots,\z_m) \, := \,
\Bigl( \,
\mathop{\prod}\limits_{1 \, \leqslant \, k \, < \, l \, \leqslant \, m}
(\z_{kl}^2)^{N_{kl}}
\Bigr) \,
\phi_{i_1} (\z_1) \dotsm \phi_{i_m} (\z_m) \,,
\]
where $\z_{kl} = \z_k - \z_l$ and
$N_{kl}$ are the integers fulfilling the locality
condition~(\ref{eq3.8n2}) for $\phi_{i_k}$ and $\phi_{i_l}$.
We then claim that all coefficients of
$A (\z_1,\dots,\z_m) \, \vac$ belong to
the image of~(\ref{eq3.12n2}).
To prove this, first note that by \ref{pr3.1n2}(e) 
$A$ is a translation invariant $m$--field
that is local with respect to $\phi_{i} (\z)$
for all $i \in \mathcal{I}$.
Then all coefficients $\psi_i(\z)$ in the operator expansion of $A$
are contained in $\mathcal{F}$ (see \eqref{eq3.10n2}).
It follows from \ref{pr3.1n2}(c) that for $v=\vac$
the right--hand side of \eqref{eq3.10n2} is simply
the Taylor expansion of
\[
A (\z+\w_1,\dots,\z+\w_{m-1},\z) \, \vac
\, \in \,
V \llbracket \z, \w_1,\dots,\w_{m-1} \rrbracket \,.
\]
Then it is clear that all coefficients of
$A (\z_1,\dots,\z_m) \, \vac$ belong to
the image of~(\ref{eq3.12n2}).

On the other hand, iterating (\ref{eq3.8n3}) we obtain that
(cf.\ \eqref{suc-loc}):
\beq\label{eq3.16n2}
\phi_{i_1} (\z_1) \dotsm \phi_{i_m} (\z_m) \, \vac \, \in \,
V \llbracket \z_1 \rrbracket_{\z^2_1} \dotsm
\llbracket \z_m \rrbracket_{\z^2_m} \,.
\eeq
The right--hand side of \eqref{eq3.16n2} is a module
over the algebra
$\CC \llbracket \z_1 \rrbracket_{\z^2_1} \dotsm
\llbracket \z_m \rrbracket_{\z^2_m}$,
in which the polynomial
\(\mathop{\prod}\limits_{k \, < \, l}
(\z_{kl}^2)^{N_{kl}}\) 
is invertible:
its inverse is given by applying the expansion
$\mathop{\prod}\limits_{k<l}$ $\iota_{\z_k,\z_l}$
(see Sect.~\ref{se1.2}).
Therefore,
\[
\phi_{i_1} (\z_1) \dotsm \phi_{i_m} (\z_m) \, \vac \, = \,
\Bigl( \,
\mathop{\prod}\limits_{1 \, \leqslant \, k \, < \, l \, \leqslant \, m}
\iota_{\z_k,\z_l} \, (\z_{kl}^2)^{-N_{kl}}
\Bigr) \,
A (\z_1,\dots,\z_m) \, \vac \,.
\]

This implies that every coefficient of (\ref{eq3.16n2}) can be expressed as
a linear combination of coefficients of
$A (\z_1,\dots,\z_m) \, \vac$, and hence belongs to
the image of map~(\ref{eq3.12n2}).
But by completeness the coefficients of (\ref{eq3.16n2}) span $V$;
therefore, (\ref{eq3.12n2}) is surjective.\end{Proof}

\skp

\begin{Corollary}\label{cr3.3n2}
Let\/ $\chi (\z)$ be a translation invariant field, which is
local with respect to a translation invariant, local and complete
system of fields\/ $\{\phi_{i} (\z) \}$.
Then\/ $\chi (\z)$ is a local field and\/
$\chi (\z) = Y (a,\z)$ for\/
$a = \chi (\z) \vac |_{\z = \zero}$.
\end{Corollary}

\skp

\ref{th3.2n2} leads naturally to the following definition
\cite{N05}.

\skp

\begin{Definition}\label{df3.1n2}
A \textbf{vertex algebra} $V$ over $\CC^D$ is a superspace $V$
endowed with{\rm:}

\medskip

{\rm (a)}
an action of mutually \emph{commuting,} \emph{even} endomorphisms
$T_1,\dots,T_D$ (\textbf{trans\-la\-tion endomorphisms}),

\medskip

{\rm (b)}
an \emph{even} vector $\vac$ (\textbf{vacuum}) such that
$T_1 \, \vac = \dots = T_D \, \vac = 0$,

\medskip

{\rm (c)}
a \emph{parity preserving} linear map
(\textbf{state--field correspondence})
\[
V \to (\End V) \llbracket \z,1/\z^2 \rrbracket \,, \quad
a \mapsto Y (a,\z) \,,
\]
such that
$\{ Y (a,\z) \}_{a \in V}$ is a
\emph{translation invariant}, \emph{local} system of \emph{fields} and
$a = Y (a,\z) \, \vac \, |_{\z = \zero}$ for
all $a \in V$.
\end{Definition}

\skp

\begin{Corollary}\label{rm3.1n3}
Every translation invariant, local and complete
system of fields $\{\phi_{i} (\z) \}$
generates on $V$ a unique structure of a vertex algebra.
\end{Corollary}

\skp

Note that when 
$D=1$, \ref{df3.1n2}
is equivalent to the definition of a usual
(chiral) vertex algebra (see e.g.\ \cite{K,LL}).
We will use the notation
\beq\label{eq3.20n2}
a (\z) \, \equiv \, Y (a,\z)
\eeq
as it is customary in the usual $D=1$ theory of vertex algebras.
For a vertex algebra $V$ and elements $a,b \in V$, we denote by
$N (a,b)$ the smallest non-negative integer
fulfilling the locality condition
for $a(\z)$ and $b(\z)$, i.e.,
\beq\label{eq3.19n2}
N (a,b) \, := \,
\mathop{\min}
\Bigl\{ N \in \NNN \, \big| \,
((\z - \w)^2)^N \,
[a(\z),b(\w)] \, = \, 0
\Bigr\} \,.
\eeq

\skp

\begin{Remark}\label{rm3.3n4}
One can introduce a more general notion of vertex algebra that
involves non-integral powers of $\z^2$ in the definition of
a field and in the notion of locality.
For $D=1$ this would correspond to the
``generalized vertex algebras'' of \cite{FFR,DL,M},
as explained in \cite{BK2}.
For $D=2$, a related notion was introduced in \cite{KO}.
\end{Remark}

\subsection{Examples of Vertex Algebras}\label{se-ex}\bsec
For completeness, in this subsection we present two
simple examples of vertex algebras.
We refer to \cite{N05,NT05,BN,BNRT} for additional examples.

Our first example shows that the notion of a
vertex algebra includes as a special case that of a
(super)commutative associative algebra (cf.\ \cite{B1}).
We call a vertex algebra $V$ over $\CC^D$ \textbf{holomorphic} if
$Y (a,\z)$ $\in$ $(\End \, V) \llbracket \z \rrbracket$
for all $a$ $\in$ $V$.
The following statement is a straightforward generalization of
the corresponding one in the case $D=1$ 
(see \cite{B1,K,LL}).

\skp

\begin{Proposition}\label{Pr4.1n1}
{\rm (a)}
If\/ $V$ is a holomorphic vertex algebra over\/ $\CC^D$, then\/
$a \star b$ $:=$ $Y (a,\z) \, b \, |_{\z = \zero}$ defines on\/
$V$ the structure of a super-commutative associative algebra 
with a unit\/ $\vac$ and even derivations $T_1,\dots,T_D$.

\medskip

{\rm (b)}
Conversely, given a super-commutative associative algebra $V$
with a product $\star$, a unit\/ $\vac$,
and with mutually commuting even derivations
$T_1,\dots,T_D$, then
$Y (a,$ $\z) \, b$ $:=$ $(e^{\z \spr \T} a) \star b$
defines the structure of a holomorphic vertex algebra on~$V$.
%
\end{Proposition}

\skp


Our second example is the vertex algebra generated by a
\textbf{harmonic scalar free field} $\varphi (\z)$
in \emph{even} space--time dimensions $D>2$ (see \cite[Sect.~5]{N05}
for a more general construction).
%
%
%
Consider a new formal variable $\p := (\pp_1,\dots,\pp_D)$ 
and introduce the vector spaces
\beq\label{eq4.3n2}
P \, := \, \CC [\p] / \p^2 \CC [\p] \cong \CC[\p]^\harm
\,, \qquad
V \, := \, \CC [P] \,.
\eeq
Here $V$ is defined as the algebra of all polynomials of the elements
of $P$, i.e., it is the symmetric algebra of $P$.
We will denote by $[f(\p)]$ the equivalence class of a polynomial
$f(\p) \in \CC [\p]$ in $P$. Then $V$ is the 
linear span over $\CC$ of all monomials of the form
$[f_1 (\p)] \cdots [f_n (\p)]$
(note that $[f_1 (\p)] [f_2 (\p)] \neq [f_1 (\p) f_2 (\p)]$ in $V$).

We define an $(\End V)$--valued formal distribution
$\varphi (\z)$ by the formula
\beq\label{eq4.4n2}
\begin{split}
&
\varphi (\z) \, [f_1 (\p)] \cdots [f_n (\p)] \, := \,
\mathop{\sum}\limits_{k \, = \, 0}^{\infty}
\, \frac{1}{k!} \, [(\z \spr \p)^k] \, [f_1 (\p)] \cdots [f_n (\p)]
\\ & \qquad + \,
\mathop{\sum}\limits_{l \, = \, 1}^{n}
\, f_l (-\Di_{\z}) (\z^2)^{-\frac{D}{2}+1} \ [f_1 (\p)] \cdots
[\widehat{f_l (\p)}] \cdots [f_n (\p)] \,,
\end{split}
\eeq
where a hat over a term means that it is omitted in the product.
This does not depend on the choice of representatives $f_l (\p)$ because
$\Di_{\z}^2 \, (\z^2)^{-\frac{D}{2}+1} = 0$.
Since the right--hand side of \eqref{eq4.4n2} contains only
finitely many negative powers of $\z^2$, it follows that 
$\varphi (\z)$ is a field.
After a straightforward computation, one finds
\beq\label{eq4.5n2}
\varphi (\z_1) \, \varphi (\z_2)  \, [f_1 (\p)] \cdots [f_n (\p)]
\, = \,
e^{ -\z_2 \spr \Di_{\z_1} }
(\z_1^2)^{-\frac{D}{2}+1}
\ [f_1 (\p)] \cdots [f_n (\p)]
\, + \, \cdots \;
%
\eeq
where the remaining terms are symmetric under the exchange of 
$\z_1$ and $\z_2$.
Hence, 
\beq\label{eq4.6n2}
[ \varphi (\z_1), \varphi (\z_2) ] \, = \,
\bigl( \iota_{\z_1,\z_2} (\z_{12}^2)^{-\frac{D}{2}+1}
-\iota_{\z_2,\z_1} (\z_{12}^2)^{-\frac{D}{2}+1} \bigr) \, \id_V \,,
\eeq
where $\z_{12} = \z_1 - \z_2$.
Therefore, $\varphi (\z)$ is a \emph{local field}, since
\beq\label{eq4.6n2a}
(\z_{12}^2)^{\frac{D}{2}-1} \,
[ \varphi (\z_1), \varphi (\z_2) ] \, = \, 0 \,.
\eeq

We define endomorphisms $T_1, \dots, T_D$ of $V$ by the formula
\beq\label{eq4.7n2}
T_{\al} \, [f_1 (\p)] \cdots [f_n (\p)] \, := \,
\mathop{\sum}\limits_{l \, = \, 1}^{n}
\; [\pp_{\al} \, f_l(\p)] \, [f_1 (\p)] \cdots
[\widehat{f_l (\p)}] \cdots [f_n (\p)]
\eeq
for 
$\al=1,\dots,D$.
In particular, $T_\al \vac = 0$, where $\vac$ is the constant
polynomial $1 \in V$. One can easily verify that the endomorphisms
$T_1, \dots, T_D$ commute with each other, and the field $\varphi (\z)$ is
\emph{translation invariant}. 
Let us point out that 
$\varphi (\z) \vac = [e^{\z\spr\p}] = e^{\z\spr\T}[1]$,
and $\varphi (\z) \vac |_{\z=0} = [1] \ne 1 = \vac$.

Using \eqref{eq4.4n2}, one can prove by induction on $n$ 
that every monomial $[f_1 (\p)]$ $\cdots$ $[f_n (\p)]$ is a linear
combination of the coefficients of 
$\varphi (\z_1) \dotsm \varphi (\z_n) \vac$.
Therefore, the field $\varphi (\z)$ is \emph{complete},
and one can apply 
\ref{rm3.1n3} to generate on $V$ the structure of
a vertex algebra over $\CC^D$.
Finally, we note that the field $\varphi (\z)$ is \emph{harmonic},
i.e., it satisfies the Laplace equation
$\Di_{\z}^2 \, \varphi (\z) = 0$.
This follows from \eqref{eq4.3n2}, \eqref{eq4.4n2} and the fact that
the function $(\z^2)^{-\frac{D}{2}+1}$ is harmonic.

\section{Jacobi Identity}\label{se4}
\setcntrs

This section contains the main results of the paper. We first prove
certain ``formal commutativity and associativity'' relations,
and then use them to derive our Jacobi identity. Integral versions
of the Jacobi identity are also presented. We show that together with
a partial vacuum axiom it can be taken as an equivalent definition
of the notion of vertex algebra over $\CC^D$. We derive a formula
for the commutator of two fields, and we prove that the singular 
parts of fields close a substructure under the commutator.

\subsection{Commutativity and Associativity}\label{se3.3}\bsec
In this subsection we will extend the ``associativity''
of \cite[Theorem 4.3]{N05} by giving a connection between the
degrees of the poles in a product $a (\z) \, b (\w)$ of two fields and
the integers $N (a,b)$ introduced in (\ref{eq3.19n2}).
In the case of usual $D=1$ vertex algebras our results agree
with the ``formal commutativity and associativity'' of 
\cite{FB,LL}.

\skp

\begin{Theorem}\label{th3.5n2}
{\rm (``Formal commutativity and associativity.'')} \,
Let $V$ be a vertex algebra, and let $a,b,c \in V$, where
$a$ and $b$ have parities $p_a$ and $p_b$, respectively.
Then there exists a localized formal series
\beq\label{eq3.21n2}
\mathcal{Y}_{a,b,c} (\z_1,\z_2) \, = \,
\frac{\psi_{a,b,c} (\z_1,\z_2)}{
(\z^2_1)^{N (a,c)} \, (\z^2_2)^{N (b,c)} \,
(\z^2_{12})^{N (a,b)}} \;,
\eeq
where
\[
\psi_{a,b,c} (\z_1,\z_2) \in
V \llbracket \z_1,\z_2 \rrbracket \,, \quad
\z_{12} = \z_1 - \z_2 \,,
\]
such that
\beq\label{eq3.22n2}
\mathcal{Y}_{a,b,c} (\z_1,\z_2) \, = \,
(-1)^{p_a p_b} \, \mathcal{Y}_{b,a,c} (\z_2,\z_1)
\eeq
and
\begin{align}
\label{eq3.23n2}
a (\z_1) \hspace{2pt} b (\z_2) \hspace{2pt} c \, &=
\iota_{\z_1,\z_2} \,
\mathcal{Y}_{a,b,c} (\z_1,\z_2) \,,
\raisebox{10pt}{} \\ \label{eq3.24n2}
b (\z_2) \hspace{2pt} a (\z_1) \hspace{2pt} c \, &=
(-1)^{p_a p_b}
\iota_{\z_2,\z_1} \,
\mathcal{Y}_{a,b,c} (\z_1,\z_2) \,,
\raisebox{10pt}{} \\ \label{eq3.25n2}
(a (\z_3) \hspace{2pt} b) (\z_2) \hspace{2pt} c \, &=
\iota_{\z_2,\z_3} \,
\mathcal{Y}_{a,b,c} (\z_2+\z_3,\z_2) \,.
\end{align}
\end{Theorem}

\skp

\begin{Proof}
Eqs.~(\ref{eq3.22n2})--(\ref{eq3.25n2}) follow from
Proposition~4.2 and Theorem~4.3 of \cite{N05}.
The explicit form~(\ref{eq3.21n2}) of
$\mathcal{Y}_{a,b,c} (\z_1,\z_2)
\in V \llbracket \z_1,\z_2 \rrbracket_{\z^2_1 \z^2_2 \z^2_{12}}$
follows from Eq.~(\ref{eq3.19n2}) and 
the following lemma, which
provides another description of the numbers $N(a,b)$.\end{Proof}

\skp

\begin{Lemma}\label{lm3.6n2}
For any two elements $a$ and $b$ in a vertex algebra $V$, we have
\beq\label{eq3.26n2}
N(a,b) \, = \,
\mathop{\min}
\Bigl\{
N \in \NNN \, \big| \,
(\z^2)^N \,
a(\z) \, b \in V \llbracket \z \rrbracket
\Bigr\} \,.
\eeq
\end{Lemma}

\skp

\begin{Proof}
Denote the right--hand side of (\ref{eq3.26n2}) by $N'(a,b)$,
and consider the formal distribution
\begin{equation*}
F_{a,b} (\z_1,\z_2) \, := \,
(\z^2_{12})^{N (a,b)} \, a(\z_1) \, b (\z_2) \,.
\end{equation*}
Due to \ref{pr3.1n2}(e), (c),
$F_{a,b}$ is a translation invariant, bilocal field and
$F_{a,b} (\z_1,$ $\z_2)$ $\vac \in V \llbracket \z_1,\z_2 \rrbracket$.
Therefore, setting $\z_2 \hspace{-1pt} = \hspace{-1pt} \zero$ we find from
$b (\z_2) \vac |_{\z_2 = \zero} = b$ that
$N' (a,b)$ $\leqslant$ $N(a,b)$.

Consider now the formal distribution
\begin{equation*}
F_{a,b}' (\z_1,\z_2) \, := \,
(\z^2_{12})^{N' (a,b)} \, a(\z_1) \, b (\z_2) \,.
\end{equation*}
As in the proof of \ref{pr3.1n2}(e), 
$F_{a,b}'$ is translation invariant and
local (as a formal distribution) with respect to all fields $c (\z)$
for $c \in V$.
It follows from translation invariance that
\beq\label{eq3.29n2}
F_{a,b}' (\z_1,\z_2) \, \vac \, = \,
e^{\z_2 \spr \T}
\left(\raisebox{10pt}{\hspace{-2pt}}\right.
(\z^2_{12})^{N' (a,b)} \, a (\z_{12}) \, b
\left.\raisebox{10pt}{\hspace{-2pt}}\right)
\in V \llbracket \z_{12},\z_2 \rrbracket
\, \equiv \, V \llbracket \z_1,\z_2 \rrbracket \,.
\eeq
On the other hand, by locality
(assuming that $c \in V$ has a fixed parity $p_c$)
we get
\begin{equation*}\label{eq3.30n2}
\begin{split}
\bigl( (\z_1 &- \w)^2 \, (\z_2 - \w)^2 \bigr)^N
F_{a,b}' (\z_1,\z_2) \, c (\w) \, \vac
\\ &= \,
\bigl( (\z_1 - \w)^2 \, (\z_2 - \w)^2 \bigr)^N
(-1)^{(p_a + p_b) p_c}
c (\w) \, F_{a,b}' (\z_1,\z_2) \, \vac
\end{split}
\end{equation*}
for $N \gg 0$.
It follows from Eq.~(\ref{eq3.29n2}) and \ref{pr3.1n2} that
both sides of the above equation
belong to the intersection
of $V \llbracket \z_1,$ $1/\z^2_1;$ $\z_2,1/\z^2_2 \rrbracket$
$\llbracket \w \rrbracket$ and
$V \llbracket \w \rrbracket_{\w^2}$
$\llbracket \z_1,\z_2 \rrbracket$.
But the latter space is exactly $V \llbracket \z_1,\z_2,\w \rrbracket$.
Therefore,
\begin{equation*}
\bigl( (\z_1 - \w)^2 \, (\z_2 - \w)^2 \bigr)^N
F_{a,b}' (\z_1,\z_2) \, c (\w) \, \vac
\in V \llbracket \z_1,\z_2,\w \rrbracket \,,
\end{equation*}
and setting $\w = \zero$ we find that $F_{a,b}'$ is a field.

In the same way one proves that
\[
F_{b,a}' (\z_1,\z_2) \, := \,
(\z^2_{12})^{N' (a,b)} \, b (\z_2) \, a (\z_1)
\]
is a field;
note that $N'(a,b)$ $=$ $N'(b,a)$ because of the 
``quasisymmetry'' relation 
\beq\label{eq-qs}
a(\z) \, b \, = \, (-1)^{p_a p_b} \, 
e^{\z \spr \T} \bigl( b(-\z) \, a \bigr) \,, 
\qquad a,b \in V 
\eeq
(see \cite[Proposition~4.4]{N05}).
Locality for $a (\z)$ and $b (\z)$ implies that
\[
(\z^2_{12})^N \,
\bigl( F_{a,b}' (\z_1,\z_2) -
(-1)^{p_a p_b} \,
F_{b,a}' (\z_1,\z_2) \bigr) \, c \, = \, 0 \,,
\qquad N \gg 0 \,.
\]
On the other hand, since $F_{a,b}'$ and $F_{b,a}'$ are fields,
\[
\bigl( F_{a,b}' (\z_1, \z_2) - (-1)^{p_a p_b}
F_{b,a}' (\z_1, \z_2) \bigr) \, c
\in V \llbracket \z_1, \z_2 \rrbracket_{\z^2_1\z^2_2} \,,
\]
which is a $\CC [ \z_1,\z_2 ]_{\z^2_1\z^2_2}$--module
with \textit{no zero divisors}.
Hence, $F_{a,b}' = (-1)^{p_a p_b} F'_{b,a}$
and $N (a,b) \leqslant N'(a,b)$.\end{Proof}

\skp

{}From the proof of \ref{lm3.6n2} and from \ref{pr3.1n2},
we deduce the following corollary.

\skp

\begin{Corollary}\label{rm3.1n2}
Let $V$ be a vertex algebra, and let $A (\z_1,\dots,\z_m)$ be an
$(\End V)$--valued formal distribution,
which is translation invariant
and local with respect to all fields $Y (c,\z)$ $(c \in V)$.
Then $A$ is an $m$--field if and only if
$A (\z_1, \dots, \z_m) \vac \in
V \llbracket \z_1, \dots, \z_n \rrbracket$.
\end{Corollary}

\skp

Next, we will derive from \ref{th3.5n2} an ``associativity'' property,
which generalizes Eqs. (4.2) and (7.3) from \cite{BK}
(see also \cite{DL,LL}).

\skp

\begin{Proposition}\label{cr3.6n3}
{\rm (``Associativity.'')} \,
For every three elements $a,b,c$ in a vertex algebra $V$
and for $L \geqslant N(a,c)$, we have{\rm:}
\beq\label{eq3.32n3}
\bigl( (\z+\w)^2 \bigr)^L \, 
\bigl( a (\z) \, b \bigr) (\w) \, c
\, = \,
\bigl( (\z+\w)^2 \bigr)^L \, \hiota_{\z,\w} \, 
a(\z+\w) \, b(\w) \, c
\,,
\eeq
\beq\label{assoc2}
(\z^2)^L \, a(\z) \, b(\w) \, c
\, = \,
\Bigl[
\bigl( (\u+\z-\w)^2 \bigr)^L \,
\hiota_{\z,\w} \,
\bigl( a (\z-\w) \, b \bigr) (\u) \, c
\, \Bigr]_{\hspace{1pt}\u = \w}
\,,
\eeq
where the expression under the substitution in the
right--hand side of \eqref{assoc2} belongs to\/
$\iota_{\z,\w} \, V \llbracket \z,\w,\u \rrbracket_{(\z-\w)^2 \com \u^2}$
and setting\/ $\u=\w$ makes sense.
\end{Proposition}

\skp

\begin{Proof}
We can assume without loss of generality that $L=N(a,c)$.
Then, by \ref{th3.5n2},
the left--hand side of (\ref{eq3.32n3}) is equal to
\[
\Txfrac{\psi_{a,b,c} (\z+\w,\w)}{(\z^2)^{N (a,b)} (\w^2)^{N (b,c)}}
\in V \llbracket \z, \w \rrbracket_{\z^2\w^2} \,,
\]
while the right--hand side is
\[
\hiota_{\z,\w} \, \iota_{\z+\w,\w} \,
\frac{\psi_{a,b,c} (\z+\w,\w)}
{\bigl( ((\z+\w)-\w)^2 \bigr)^{N (a,b)} \bigl( \w^2 \bigr)^{N (b,c)}} \;.
\]
Then Eq.~(\ref{eq3.32n3}) follows from Taylor's formula~(\ref{eq2.30n3}).
The proof of Eq.~\eqref{assoc2} is simpler: its sides 
are equal to
\[
\iota_{\z,\w} \,
\Txfrac{\psi_{a,b,c} (\z,\w)}{\bigl((\z-\w)^2\bigr)^{N (a,b)} (\w^2)^{N (b,c)}}
\quad \text{and} \quad
\hiota_{\z,\w} \,
\frac{\psi_{a,b,c} (\u+\z-\w,\u)}
{\bigl( (\z-\w)^2 \bigr)^{N (a,b)} \bigl( \u^2 \bigr)^{N (b,c)}} \;,
\]
respectively, and obviously they become equal after the substitution $\u=\w$.\end{Proof}

\skp

\subsection{Jacobi Identity}\label{se-jac}\bsec
In this subsection, for any three elements in 
a vertex algebra over $\CC^D$, we derive an identity
that generalizes the Jacobi identity of \cite{FLM} 
(and the Borcherds identity of \cite{K})
for usual $D=1$ vertex algebras. 

\skp

\begin{Theorem}\label{th-jac}
{\rm (``Jacobi identity.'')} \,
Let\ $V$ be a vertex algebra, and let\/ $a,b,c \in V$, where
$a$ and $b$ have fixed parities $p_a$ and $p_b$, respectively.
Then for $L \geqslant N(a,c)$ and
for every $F (\z,\w) \in
\CC \llbracket \z,\w \rrbracket_{(\z^2)^{\RR} \com (\w^2)^{\RR} \com
((\z - \w)^2)^{\RR}}$,
we have{\rm:}
\beq\label{eq-jac1}
\begin{split}
a(& \z) \, b(\w) \, c \; \iota_{\z,\w} F(\z,\w)
\, - \,
(-1)^{p_a p_b} \, b(\w) \, a(\z) \, c \; \iota_{\w,\z} F(\z,\w)
\\ 
& = \,
(\z^2)^{-L} \Bigl[ \bigl( (\u+\z-\w)^2 \bigr)^{L} \, 
(\iota_{\z,\w} - \iota_{\w,\z}) \,
\bigl( a (\z-\w) \, b \bigr) (\u) \, c \; F(\z,\w) 
\Bigr]_{\u=\w}
\, ,
\end{split}
\eeq
where the expression under the substitution in
the right--hand side belongs to\/
$(\iota_{\z,\w} - \iota_{\w,\z} )$
$V \llbracket \z, \w,\u 
\rrbracket_{(\z^2)^\RR \com (\w^2)^\RR \com ((\z-\w)^2)^\RR \com \u^2}$
and setting\/ $\u=\w$ makes sense.
\end{Theorem}

\skp

\begin{Proof}
By the same argument as in the proof of Eq.~\eqref{assoc2} above, 
one finds separately
\beqa
\label{assoc1a}
&&
\begin{aligned}
a(\z) & \, b(\w) \, c \; \iota_{\z,\w} F(\z,\w)
\\ 
& = \,
(\z^2)^{-L} \, \Bigl[ \bigl( (\u+\z-\w)^2 \bigr)^{L} \, 
\iota_{\z,\w} \,
\bigl( a (\z-\w) \, b \bigr) (\u) \, c \; F(\z,\w) 
\Bigr]_{\u=\w}
\, , 
\end{aligned}
\\ &&
\raisebox{26pt}{}
\label{assoc2a}
\begin{aligned}
(-1 &)^{p_a p_b} \, b(\w) \, a(\z) \, c \; \iota_{\w,\z} F(\z,\w)
\\ 
& = \,
(\z^2)^{-L} \, \Bigl[ \bigl( (\u+\z-\w)^2 \bigr)^{L} \, 
\iota_{\w,\z} \,
\bigl( a (\z-\w) \, b \bigr) (\u) \, c \; F(\z,\w) 
\Bigr]_{\u=\w}
\,
\end{aligned}
\eeqa
for $L \geqslant N(a,c)$.
Taking the difference we obtain~\eqref{eq-jac1}.\end{Proof}

\skp

The main subtlety of Eq.~\eqref{eq-jac1} is that in the right--hand side
one can \emph{not} make the substitution $\u=\w$ in each of the factors
separately. It is only after we multiply them that this substitution 
makes sense. The reason is that, in contrast to the case $D=1$,
the expression 
$(\iota_{\z,\w} - \iota_{\w,\z}) \, a (\z-\w) \, b$
involves an infinite sum, and hence in general
$(\iota_{\z,\w} - \iota_{\w,\z}) \, ( a (\z-\w) \, b ) (\w) \, c $
is not well defined. 
We refer to Sect. \ref{se-jco} and \ref{se5.3}
below for additional discussion.

It is clear from the proof of Eq.~\eqref{assoc2} 
that if we multiply the right--hand sides 
of Eqs.~\eqref{assoc1a} and \eqref{assoc2a} by $(\u^2)^{M}$ for 
$M \geqslant N (b,c)$, they will become regular in $\u$ 
(i.e., not containing negative powers of $\u^2$).
Then we will be able to represent the substitution $\u=\w$ 
by Cauchy formula \eqref{eqn3nn6}.
Thus we obtain the following equivalent integral form of 
Jacobi identity~\eqref{eq-jac1}.

\skp

\begin{Corollary}\label{cr-jac}
For every elements
$a,b,c$ in a vertex algebra $V$, $a$ and $b$ having fixed parities 
$p_a$ and $p_b$, respectively,
and for every $L \geqslant N(a,c)$, $M \geqslant N(b,c)$, we have{\rm:}
\beq\label{eq-jac1a}
\begin{split}
a(\z) \, b( & \w) \, c \; \iota_{\z,\w} F(\z,\w)
\, - \,
(-1)^{p_a p_b} \, b(\w) \, a(\z) \, c \; \iota_{\w,\z} F(\z,\w)
\\ 
= \, & \Res_\u \;
(\z^2)^{-L} \, (\w^2)^{-M} \, \bigl( (\u+\z-\w)^2 \bigr)^{L} \, (\u^2)^{M}
\, \iota_{\u,\w} \bigl( (\u-\w)^2 \bigr)^{ -\frac{D}{2} }
\\ 
& \times
(\iota_{\z,\w} - \iota_{\w,\z}) \,
\Bigl(
\bigl( a (\z-\w) \, b \bigr) (\u) \, c \; F(\z,\w) 
\Bigr)
\,
\end{split}
\eeq
for
$F (\z,\w) \in
\CC \llbracket \z,\w \rrbracket_{(\z^2)^{\RR} \com (\w^2)^{\RR} \com
((\z - \w)^2)^{\RR}}$.
\end{Corollary}

\skp

\begin{Proof}
It remains to note that the right--hand side of \eqref{eq-jac1a}
makes sense.
Indeed, the product of the Cauchy kernel 
$\iota_{\u,\w} ((\u$ $-$ $\w)^2)^{-\frac{D}{2}}$
and the third line in \eqref{eq-jac1a} is well defined in the space
\(
(\iota_{\z,\w} - \iota_{\w,\z}) \,
V\llbracket \u \rrbracket_{\u^2} 
\llbracket \z,\w \rrbracket_{(\z^2)^{\RR} \com (\w^2)^{\RR} \com ((\z-\w)^2)^{\RR}}
.
\)\end{Proof}

\skp

\begin{Remark}\label{rm3.2n3}
One can give an alternative proof of \ref{th-jac}
by using ``associativity'' relation \eqref{eq3.32n3},
``quasisymmetry'' relation \eqref{eq-qs}, 
and generalizing the arguments of \cite[Sect.~7]{BK} 
to the case of arbitrary $D$ (see also \cite{FHL,L1,LL}).
With obvious modifications, Eq.~\eqref{eq-jac1} remains
valid for generalized vertex algebras (see \ref{rm3.3n4} 
and~\cite{FFR,DL,M,BK2}).
\end{Remark}

\skp

We will show in Sect.~\ref{se-jco} below that Jacobi identity \eqref{eq-jac1},
together with a partial vacuum axiom, can be taken as an equivalent definition
of vertex algebra over $\CC^D$.

\subsection{Integral Borcherds Formula}\label{se3.4}\bsec
In this subsection we will derive an integral version of 
Jacobi identity \eqref{eq-jac1}, which in particular gives
a formula for the commutator of modes that generalizes the
Borcherds commutator formula from \cite{B1} 
(see also~\cite{FLM,FHL,K,LL}).

Let us introduce the following additive subgroup of $\RR$,
\beq\label{eq3.33n5}
\ZZP \, := \, \ZZ + \tfrac{D}{2} \ZZ
\, = \,
\begin{cases}
\ZZ \,, \;\;\qquad\text{if $D$ is even;} \\
\tfrac12 \ZZ \,, \qquad\text{if $D$ is odd.}
\end{cases}
\eeq
As a consequence of \ref{pr2.3n3},
every $(\End V)$--valued formal distribution
$\phi (\z)$ $\in$ $(\End V) \llbracket \z,$ $1/z^2 \rrbracket$
can be considered as a linear map
\beq\label{eq3.33n2}
\CC [\z]_{(\z^2)^{\ZZP}} \to \End V \,, \qquad
f (\z) \mapsto \Res_{\z} \, \phi (\z) f (\z) \,.
\eeq
Thus $\CC [\z]_{(\z^2)^{\ZZP}}$ plays the role of a vector space
of \textit{test functions}, and for even space--time dimension $D$
it is exactly
$\CC [\z]_{\z^2} \equiv \CC [\z]_{(\z^2)^{\ZZ}}$.
According to Eq.~(\ref{fd12}), the modes of $\phi (\z)$
can be obtained by integrating $\phi (\z)$ (with $\Res_\z$)
against appropriate test functions.

Now let us take
$\Res_\z \, \Res_\w$ of both sides
of Eqs.~\eqref{assoc1a} and \eqref{assoc2a} for
$F (\z,\w) \in
\CC \llbracket \z,\w \rrbracket_{(\z^2)^{\ZZP} \com (\w^2)^{\ZZP} \com
((\z - \w)^2)^{\ZZP}}$,
and represent the substitution $\u=\w$ as $\Res_\u$ as done in
Eq.~\eqref{eq-jac1a}.
We are going to rewrite the resulting identities
in the form
\beqa\label{eq3.36n2}
&&
\begin{aligned}
&\Res_{\z} \, \Res_{\w} \, a(\z) \, b(\w) \, c \ \iota_{\z,\w} \, F(\z,\w)
\\ & \hspace{25pt} = \,
\Res_{\z} \, \Res_{\w} \,
\KK_{L,M}^{\hspace{1pt}+} (\z,\w;F) \, (a (\z) \, b) \hspace{1pt} (\w) \, c \,,
\end{aligned}
\\ \raisebox{10pt}{} \label{eq3.37n2} &&
\begin{aligned}
&\Res_{\z} \, \Res_{\w} \, b(\w) \, a(\z) \, c \ \iota_{\w,\z} \, F(\z,\w)
\\ & \hspace{25pt} = \,
(-1)^{p_a p_b}
\Res_{\z} \, \Res_{\w} \,
\KK_{L,M}^{\hspace{1pt}-} (\z,\w;F) \, (a (-\z) \, b) \hspace{1pt} (\w) \, c
\,,
\end{aligned}
\eeqa
where $\KK_{L,M}^\pm$ are to be determined. To arrive at the above
formulas, we will use translation invariance of the residue
to replace $\z$ with $\z+\w$ in \eqref{assoc1a} and $\w$ with $\w+\z$
in \eqref{assoc2a}. More precisely, we have the following lemma.

\skp

\begin{Lemma}\label{lm3.9n2}
For every $G (\z,\w)$ $\in$
\(V \llbracket \z,\w \rrbracket_{(\z^2)^{\ZZP} \com (\w^2)^{\ZZP} \com
((\z - \w)^2)^{\ZZP}}\)
we have
\beq\label{eq3.43n2}
\Res_{\z} \, \Res_{\w} \,
\iota_{\z,\w} \, G (\z,\w)
\, = \,
\Res_{\z} \, \Res_{\w} \,
\iota_{\z,\w} \,
G (\z \hspace{-1pt} + \hspace{-1pt} \w,\w) \,.
\eeq
\end{Lemma}

\begin{Proof}
Translation invariance of the residue
(see (\ref{eqn3n.5})) implies the identity
\[ 
\Res_{\z} \, \Res_{\w} \,
\iota_{\z,\w} \, G (\z,\w)
\, = \,
\Res_{\z} \, \Res_{\w} \,
\hiota_{\z,\u} \, \iota_{\z+\u,\w} \,
G (\z \hspace{-1pt} + \hspace{-1pt} \u,\w) \,.
\]
Then since the expression under the residue in
the right--hand side belongs to the space
$V \llbracket \z \rrbracket_{(\z^2)^{\ZZP}}
\llbracket \w,\u \rrbracket_{(\w^2)^{\ZZP}}$,
we can set there $\u = \w$.
But 
\[
\hiota_{\z,\w} \, \iota_{\z+\w,\w} \,
G (\z \hspace{-1pt} + \hspace{-1pt} \w,\w) 
=
\iota_{\z,\w} \, G (\z \hspace{-1pt} + \hspace{-1pt} \w,\w)
\]
by ``Taylor formula''~(\ref{eq2.30n3}).\end{Proof}

\skp

Applying \ref{lm3.9n2} to the right--hand side of \eqref{assoc1a},
we obtain \eqref{eq3.36n2} with
\beq\label{eq3.34n2}
\begin{split}
& \hspace{15pt}
\KK_{L,M}^{\hspace{1pt}+} (\z,\w; F) \, := \,
\Res_{\u} \
\iota_{\z,\u} \ F (\u \hspace{-1pt} + \hspace{-1pt} \z,\u)
\left(\raisebox{9pt}{\hspace{-2.5pt}}\right.
(\u \hspace{-1pt} + \hspace{-1pt} \z)^2
\left.\raisebox{9pt}{\hspace{-2.5pt}}\right)^{-L}
(\u^2)^{-M}
\\
& \hspace{15pt} \hspace{25pt} \times \
\iota_{\w,\u}
\left(\raisebox{9pt}{\hspace{-2.5pt}}\right.
(\w \hspace{-1pt} - \hspace{-1pt} \u)^2
\left.\raisebox{9pt}{\hspace{-2.5pt}}\right)^{-\frac{D}{2}}
\left(\raisebox{9pt}{\hspace{-2.5pt}}\right.
(\z \hspace{-1pt} + \hspace{-1pt} \w)^2
\left.\raisebox{9pt}{\hspace{-2.5pt}}\right)^{L}
(\w^2)^{M} \,.
\end{split}
\eeq
Similarly, after a renaming of the variables,
\eqref{assoc2a} leads to \eqref{eq3.37n2} with
\beq\label{eq3.35n2}
\begin{split}
& \hspace{15pt}
\KK_{L,M}^{\hspace{1pt}-} (\z,\w; F) \, := \,
\Res_{\u} \
\iota_{\z,\u} \ F (\u,\u \hspace{-1pt} + \hspace{-1pt} \z)
\left(\raisebox{9pt}{\hspace{-2.5pt}}\right.
(\u \hspace{-1pt} + \hspace{-1pt} \z)^2
\left.\raisebox{9pt}{\hspace{-2.5pt}}\right)^{-M}
(\u^2)^{-L}
\\ & \hspace{15pt} \hspace{25pt} \times \
\iota_{\w,\u \hspace{-1pt} + \hspace{-1pt} \z}
\left(\raisebox{9pt}{\hspace{-2.5pt}}\right.
(\w-(\u \hspace{-1pt} + \hspace{-1pt} \z))^2
\left.\raisebox{9pt}{\hspace{-2.5pt}}\right)^{-\frac{D}{2}}
\left(\raisebox{9pt}{\hspace{-2.5pt}}\right.
(\w \hspace{-1pt} - \hspace{-1pt} \z)^2
\left.\raisebox{9pt}{\hspace{-2.5pt}}\right)^{L}
(\w^2)^{M}
\, .
\end{split}
\eeq
Notice that the expressions after $\Res_\u$
in the right--hand sides of (\ref{eq3.34n2}) and
(\ref{eq3.35n2}) are well-defined elements of
\(\CC \llbracket \w \rrbracket_{(\w^2)^{\ZZP}}
\llbracket \z \rrbracket_{(\z^2)^{\ZZP}}
\llbracket \u \rrbracket_{(\u^2)^{\ZZP}}\),
and in fact the former belongs to
\(\CC \llbracket \z,\w \rrbracket_{(\z^2)^{\ZZP} \com (\w^2)^{\ZZP}}
\llbracket \u \rrbracket_{(\u^2)^{\ZZP}}\).
Then (\ref{eq3.34n2}) and the formula
\beq\label{eq3.35n3}
\KK_{L,M}^{\hspace{1pt}-} (\z,\w; F) \, = \,
\io_{\w,\z} \, \KK_{L,M}^{\hspace{1pt}+} (\z,\w-\z; F^\op) \,,
\qquad 
F^\op (\z,\w) := F (\w,\z) 
\eeq
imply that
\beq\label{eq3.36n3}
\KK_{L,M}^{\hspace{1pt}+} (\z,\w; F) \, \in \,
\CC \llbracket \z,\w \rrbracket_{(\z^2)^{\ZZP} \com (\w^2)^{\ZZP}}
\, , \quad
\KK_{L,M}^{\hspace{1pt}-} (\z,\w; F) \, \in \,
\CC \llbracket \w \rrbracket_{(\w^2)^{\ZZP}}
\llbracket \z \rrbracket_{(\z^2)^{\ZZP}} \, .
\eeq

Taking the difference of Eqs.~\eqref{eq3.36n2} and \eqref{eq3.37n2},
and using \eqref{eq-z3},
we obtain the following result.

\skp

\begin{Theorem}\label{th-jac3}
With the above notation, in any vertex algebra, we have{\rm:}
\beq\label{eq3.36n4}
\begin{split}
\Res_{\z} \, & \Res_{\w} \, a(\z) \, b(\w) \, c \ \iota_{\z,\w} \, F(\z,\w)
\\
&- (-1)^{p_a p_b}
\Res_{\z} \, \Res_{\w} \, b(\w) \, a(\z) \, c \ \iota_{\w,\z} \, F(\z,\w)
\\
&= \, \Res_{\z} \, \Res_{\w} \,
\KK_{L,M} (\z,\w;F) \, (a (\z) \, b) \hspace{1pt} (\w) \, c \,,
\end{split}
\eeq
where
\beq\label{eq3.36n5}
\KK_{L,M} (\z,\w;F)
:= \KK_{L,M}^{\hspace{1pt}+} (\z,\w;F)
- \KK_{L,M}^{\hspace{1pt}-} (-\z,\w;F) \,.
\eeq
\end{Theorem}

\skp

In particular, when $F(\z,\w) = f(\z) \, g(\w)$ is a product
of two test functions, Eq.~\eqref{eq3.36n4} gives a formula for
the commutator of modes, generalizing the Borcherds formula.

\subsection{The Jacobi Identity As Alternative Axiom.
            The Case $D=1$}\label{se-jco}\bsec
In this subsection we derive some consequences of 
our Jacobi identity \eqref{eq-jac1}. We prove that together with
a partial vacuum axiom it can be taken as an equivalent definition
of vertex algebra over $\CC^D$. We also show that for $D=1$ it
reduces to (an equivalent form of) the Jacobi identity of~\cite{FLM}. 

First of all, it is clear from the definitions that if
$(\z^2)^N \, a(\z) b \in V \llbracket \z \rrbracket$,
then 
$(\iota_{\z,\w} - \iota_{\w,\z}) \, a (\z-\w) \, b \; ((\z-\w)^2)^N = 0$.
Therefore, Jacobi identity \eqref{eq-jac1} implies locality.
Our next step is to show that it also implies ``associativity.''

\skp

\begin{Lemma}\label{lm-jco1}
Let $V$ be a vector space, let $c$ be an element of\/ $V$, and let\/
$a(\z)$, $b(\w)$ be two fields on\/ $V$. Assume that Eq.~\eqref{eq-jac1}
holds for some fixed\/ $p_a$, $p_b$, $L$ and for all\/ 
$F (\z,\w) \in
\CC \llbracket \z,\w \rrbracket_{\z^2 \com \w^2 \com 
((\z - \w)^2)^\ZZP}$ $($see \eqref{eq3.33n5}$)$.
Then Eq.~\eqref{eq3.32n3} holds for some $L' \geqslant L$.
\end{Lemma}

\skp

\begin{Proof}
Let $L' \geqslant L$ be large enough so that 
$(\z^2)^{L'} \, a(\z) \, c \in V \llbracket \z \rrbracket$.
Obviously, if Eq.~\eqref{eq-jac1} holds for some $L$ then it holds
for all $L' \geqslant L$, so let us just assume $L'=L$.
Applying $\Res_\z$ to both sides of \eqref{eq-jac1} with
$F (\z,\w) = (\z^2)^L \, f(\z-\w)$, where 
$f(\z) \in \CC \llbracket \z \rrbracket_{(\z^2)^{\ZZP}}$ is an 
arbitrary test function, we obtain:
\begin{align*}
\Res_\z \, & (\z^2)^L \, a(\z) \, b(\w) \, c  \; \iota_{\z,\w} f(\z-\w)
\\
\, & = \,
\Res_\z \, 
\Bigl[ \bigl( (\u+\z-\w)^2 \bigr)^{L} \, 
\iota_{\z,\w} \,
\bigl( a (\z-\w) \, b \bigr) (\u) \, c \; f(\z-\w)
\Bigr]_{\u=\w}
\,.
\end{align*}
Now using the translation invariance of the residue (see \eqref{eqn3n.5}),
we get
\begin{align*}
\Res_\z \, & \bigl( (\z+\w)^2 \bigr)^L \, 
\iota_{\z,\w} \, a(\z+\w) \, b(\w) \, c  \, f(\z)
\\
\, & = \,
\Bigl[ \Res_\z \, \bigl( (\u+\z)^2 \bigr)^{L} \, 
\bigl( a (\z) \, b \bigr) (\u) \, c \; f(\z)
\Bigr]_{\u=\w}
\,.
\end{align*}
After the substitution $\u=\w$, this gives exactly Eq.~\eqref{eq3.32n3} 
(cf.\ \eqref{eq3.33n2}).\end{Proof}

\skp

Now we can prove the following statement, which shows that a vertex algebra
can be defined in terms of Jacobi identity as in \cite{FLM}
for the $D=1$ case (see also \cite{K,LL}).

\skp

\begin{Theorem}\label{th-jac2}
Let\/ $V$ be a vector superspace endowed with an even vector\/ $\vac$
and with a parity preserving linear map\/
$Y \colon V \mapsto Y(a,\z) \equiv a(\z)$ to the space of fields on $V$.
Assume that Jacobi identity~\eqref{eq-jac1}
holds for every fixed\/ $a,b,c\in V$ with parities $p_a$, $p_b$ 
of\/ $a$ and $b$, respectively, for some $L \geqslant 0$ and for all\/ 
$F (\z,\w) \in
\CC \llbracket \z,\w \rrbracket_{\z^2 \com \w^2 \com 
((\z - \w)^2)^\ZZP}$. 
Finally, let the following
``partial vacuum axiom'' be satisfied{\rm:}
\beq\label{eq-pva}
Y(\vac,\z) \, a \, = \, a \,, \qquad
\Res_\z \, (\z^2)^{-\frac{D}2} \, Y(a,\z) \, \vac \, = \, a 
\quad\text{for all}\quad a \in V \,.
\eeq
Then there exist uniquely determined mutually commuting
even endomorphisms\/ $T_1,\dots,T_D$ of\/ $V$, which make
$V$ a vertex algebra over\/ $\CC^D$.
\end{Theorem}

\skp

\begin{Proof}
We have already pointed out that locality of $a(\z)$ and $b(\w)$
follows from Jacobi identity for $F (\z,\w) = ((\z-\w)^2)^N$
with $N\gg0$. We will derive the rest of the axioms of
vertex algebra (\ref{df3.1n2}) 
from Eqs.~\eqref{eq3.32n3} and \eqref{eq-pva} (cf.\ \ref{lm-jco1}).

Putting in Eq.~\eqref{eq3.32n3} $b=\vac$ and $L\gg0$ such that
$(\z^2)^L \, a(\z) \, c \in V \llbracket \z \rrbracket$,
we obtain that 
\[
\bigl( (\z+\w)^2 \bigr)^L \, 
\bigl( a(\z) \, \vac \bigr)(\w) \, c
\in V \llbracket \z,\w \rrbracket \,.
\]
Since 
$( a(\z) \, \vac )(\w) \, c
\in V \llbracket \w \rrbracket_{\w^2} \llbracket \z\rrbracket_{\z^2}$,
it makes sense to multiply the above equation by
$\iota_{\w,\z} ( (\z+\w)^2 )^{-L}$
and get
\[
\bigl( a(\z) \, \vac \bigr)(\w) \, c
\in \iota_{\w,\z} \, V \llbracket \z,\w \rrbracket_{(\z+\w)^2} 
\subset V \llbracket \w \rrbracket_{\w^2} \llbracket \z\rrbracket \,.
\]
Then letting $c=\vac$ and using the second equality in \eqref{eq-pva}
(with respect to $\w$), we deduce from here that
$a(\z) \, \vac \in V \llbracket \z \rrbracket$ for all $a \in V$.
Thus the second equality in \eqref{eq-pva} can be restated as
$a(\z) \, \vac |_{\z=0} = a$.

We define the translation endomorphisms $T_1,\dots,T_D$ of $V$ by the formula
\[
T_\al \, a \, := \, 
\di_{z^\al} \, a(\z) \, \vac \big|_{\z=0}
\,, \qquad \al=1,\dots,D \,,
\]
which should hold if $V$ is a vertex algebra.
Then putting $c=\vac$ in Eq.~\eqref{eq3.32n3}, we deduce that
\[
T_\al \bigl( a(\z) \, b \bigr) - a(\z) \, ( T_\al \, b )
\, = \, \di_{z^\al} \, a(\z) \, b \,,
\]
while the substitution $b=\vac$ in Eq.~\eqref{eq3.32n3} implies
$(T_\al \, a)(\z) = \di_{z^\al} \, a(\z)$.
The remaining axioms (\ref{df3.1n2}(a), (b)) are then immediate.\end{Proof}

\skp

\begin{Remark}\label{rm-jac2}
With obvious modifications, \ref{th-jac} and \ref{th-jac2}
hold also for modules over vertex algebras 
(see \cite[Sect.~6]{N05} for the definition of module).
\end{Remark}

\skp

The main subtlety of Jacobi identity~\eqref{eq-jac1} is that one
can \emph{not} make the substitution $\u=\w$ in each of the factors
separately. 
However, in the next proposition we will show that
this can be done if the field $a(\z)$ has a special form.
Recall that the regular and singular parts of a formal distribution
were defined in Sect.~\ref{se-de}.

\skp

\begin{Proposition}\label{pr-jac1}
Let\/ $a,b$ be elements in a vertex algebra $V$,
with parities $p_a$ and $p_b$, respectively.
Assume that for some $n\in\ZZ$ the singular part of\/
$(\z^2)^n \, a(\z) \, b$ belongs to\/
$V [ \z ]_{\z^2}$.
Then we have{\rm:}
\beq\label{eq-jac3}
\begin{split}
a(& \z) \, b(\w) \; \iota_{\z,\w} \bigl( (\z-\w)^2 \bigr)^n
\, - \,
(-1)^{p_a p_b} \, b(\w) \, a(\z) \; \iota_{\w,\z} \bigl( (\z-\w)^2 \bigr)^n
\\ 
& = \,
(\iota_{\z,\w} - \iota_{\w,\z}) \,
\Bigl( 
\bigl( a (\z-\w) \, b \bigr) (\w) \; \bigl( (\z-\w)^2 \bigr)^n
\Bigr)
\,,
\end{split}
\eeq
and the right--hand side is well defined.
\end{Proposition}

\skp

\begin{Proof}
For an arbitrary fixed $c \in V$ and $L \gg 0$,
set $F(\z,\w) = ((\z-\w)^2)^n$ in Eq.~\eqref{eq-jac1}.
Because the regular part of $((\z-\w)^2)^n \, a(\z-\w)$ is killed by 
$\iota_{\z,\w} - \iota_{\w,\z}$,
we can replace $((\z-\w)^2)^n \, a(\z-\w)$ by its singular part
in the right--hand side of \eqref{eq-jac1}.
But by assumption the coefficients of the singular part of
$(\z^2)^n \, a(\z) \, b$ span a finite--dimensional subspace of $V$.
Therefore, in \eqref{eq-jac1} one can substitute $\u=\w$ in each factor
separately, and the right--hand side of \eqref{eq-jac3}
makes sense. After putting $\u=\w$ in the other factor in \eqref{eq-jac1}
it cancels with~$(\z^2)^L$.\end{Proof}

\skp

\begin{Remark}\label{rm-jac3}
When $D=1$, the assumption of \ref{pr-jac1} is satisfied for
every pair of elements\/ $a,b \in V$ and every $n\in\ZZ$.
In this case, the collection of identities \eqref{eq-jac3} is
equivalent to the Jacobi identity of~\cite{FLM}.
\end{Remark}

\skp

Let us note that for $D>1$ the assumption of \ref{pr-jac1} is in fact
quite restrictive. 
For $n=0$ it holds for the scalar free field $\varphi (\z)$
discussed in Sect.~\ref{se-ex} (because $\varphi (\z)$ is harmonic)
but it does \emph{not} hold for the Wick square 
$: \!\! \varphi (\z)^2 \!\! : \,$.
Furthermore, it does \emph{not} hold for $\varphi (\z)$ itself
when $n<0$. 
On the other hand, the assumption is satisfied for $n=0$ and any
``generalized free field'' (see \cite[Sect.~5]{N05}),
thus providing a version of the \emph{Wick Theorem}
(note that the element $b$ is arbitrary).

\subsection{Degree Cutoffs and Commutator Formula}\label{se5.3}\bsec
In this subsection, we derive a commutator formula, which
shows in particular
that the singular modes of fields close an algebraic structure
under the commutator. 

We have remarked at the end of the previous subsection that
the main difficulty for $D>1$ as opposed to $D=1$
is that the singular part of $a(\z) \, b$ involves an infinite sum
in general. To circumvent this problem we introduce ``degree cutoffs''
as follows. For a formal distribution 
$\FF(\z)$, written as in \eqref{fd11}, and for any $N \in \RR$,
we define the \textbf{cutoff} 
$\FF(\z)^{\leqslant N}$ 
by restricting the sums over $m$ and $\ga$ in \eqref{fd11} to indices with
$m+2\ga \leqslant N$. In other words, we restrict the sum to terms with
degrees in $\z$ less than or equal to $N$. We denote the remaining part
of $\FF(\z)$ by $\FF(\z)^{>N} := \FF(\z) - \FF(\z)^{\leqslant N}$.
In the same way, we define cutoffs $\FF(\z)_\sip^{\leqslant N}$
of the singular part of $\FF(\z)$ (see \eqref{fd17}).
Note that all these operations are commuting projections on the
space of formal distributions.

Even though
the singular part $a(\z)_\sip b$ may involve infinitely many
terms with arbitrarily high degrees in $\z$, it is important
that for fixed $N\in\ZZ$ the cutoff of the singular part
$a(\z)_\sip^{\leqslant N} b$ 
is \emph{finite}, i.e., it belongs to $V[\z]_{\z^2}$.
Then the same argument as in the proof of \ref{pr-jac1} gives
the following result.

\skp

\begin{Lemma}\label{lm-jco2}
Let\/ $a,b,c$ be elements in a vertex algebra $V$, where
$a$ and $b$ have parities $p_a$ and $p_b$, respectively.
Then for every $n,N \in\ZZ$ and every $L \geqslant N(a,c)$,
we have{\rm:}
\begin{equation*}
\begin{split}
a & (\z) \, b(\w) \; \iota_{\z,\w} \bigl( (\z-\w)^2 \bigr)^n
\, - \,
(-1)^{p_a p_b} \, b(\w) \, a(\z) \; \iota_{\w,\z} \bigl( (\z-\w)^2 \bigr)^n
\\ 
& = \,
(\iota_{\z,\w} - \iota_{\w,\z}) \,
\Bigl( 
\bigl( a (\z-\w)^{\leqslant N} \, b \bigr) (\w) \; \bigl( (\z-\w)^2 \bigr)^n
\Bigr)
\\
&+  \,
(\z^2)^{-L} \Bigl[ \bigl( (\u+\z-\w)^2 \bigr)^{L} \, 
(\iota_{\z,\w} - \iota_{\w,\z}) \,
\bigl( a (\z-\w)^{>N} \, b \bigr) (\u) \, c \; \bigl( (\z-\w)^2 \bigr)^n
\Bigr]_{\u=\w}
\,.
\end{split}
\end{equation*}
\end{Lemma}


We will now apply this lemma in the case $n=0$ when it reduces to a
formula for the commutator of the fields $a (\z)$, $b(\w)$.
Then, because of the presence of
$\iota_{\z,\w}$ $-$ $\iota_{\w,\z}$,
in the right--hand side of 
the above equation
one can replace $a (\z-\w)^{\leqslant N}$ and $a (\z-\w)^{>N}$
by their singular parts and obtain:
\beq\label{e6.1}
\begin{split}
\bigl[a(& \z) ,\, b(\w)\bigr] \, c
\, = \,
(\iota_{\z,\w} - \iota_{\w,\z})
\bigl( a (\z-\w)^{\leqslant N}_\sip \, b \bigr) (\w) 
\\
&+  \,
\Bigl[ (\z^2)^{-L} \, \bigl( (\u+\z-\w)^2 \bigr)^{L} \, 
(\iota_{\z,\w} - \iota_{\w,\z})
\bigl( a (\z-\w)^{>N}_{\sip} \, b \bigr) (\u) \, c
\Bigr]_{\u=\w}
\,.
\end{split}
\eeq
The next result shows that the singular parts of fields 
themselves close a structure with respect to the commutator.

\skp

\begin{Proposition}\label{th-sp}
For every three elements $a,b,c$ in a vertex algebra $V$ 
and for every $L \geqslant N(a,c)$, one has{\rm:}
\beq\label{e5.17}
\begin{split}
\bigl[a& (\z)_{\sip} ,\, b(\w)_{\sip}\bigr] \, c
\\
& =
\Bigl(
\Bigl[ (\z^2)^{-L} \bigl( (\u+\z-\w)^2 \bigr)^{L} \,
(\iota_{\z,\w} - \iota_{\w,\z})
\bigl( a (\z-\w)_{\sip} \, b \bigr) (\u)_{\sip} \, c
\Bigr]_{\u=\w}
\, \Bigr)_{\sip}
\end{split}
\eeq
where the outer\/ $\sip$ in the right--hand side
designates taking the singular part with respect to
both $\z$ and~$\w$.
\end{Proposition}

\skp

Note that for $D>1$ a product of two
singular terms may contain a regular part; that is why
in \eqref{e5.17} we must
include the outer projection onto the singular parts.

\skp

\begin{Proof}[Proof of \ref{th-sp}]
We will prove that \eqref{e5.17} holds for all terms with total degree 
in $\z$ and $\w$ up to $N$, for every fixed $N \in \ZZ$.
For this purpose, we consider all terms of total degree 
$\leqslant N$ in \eqref{e6.1}, 
and take the singular parts with respect to
both $\z$ and $\w$. We will consider separately the resulting two terms in
the right--hand side.

In the first term, the expansion $\io_{\w,\z}$ will not contribute 
because it produces terms regular in $\z$. Since 
$\io_{\z,\w} \bigl( a (\z-\w)_{\sip}^{\leqslant N}
 \, b \bigr) (\w)_+$ is regular in $\w$, it will not contribute either,
and we will obtain 
\[
\Bigl(
\iota_{\z,\w} 
\bigl( a (\z-\w)_{\sip}^{\leqslant N} \, b \bigr) (\w)_{\sip} \, c
\Bigr)_{\sip} \,.
\]
Reversing the above reasoning, this expression can be rewritten as
\[
\Bigl(
(\iota_{\z,\w} - \iota_{\w,\z})
\bigl( a (\z-\w)_{\sip}^{\leqslant N} \, b \bigr) (\w)_{\sip} \, c
\, \Bigr)_{\sip}
\,.
\]
Then, as in the derivation of \ref{lm-jco2}, it is equal to
\[
\Bigl(
\Bigl[ (\z^2)^{-L} \, \bigl( (\u+\z-\w)^2 \bigr)^{L} \,
(\iota_{\z,\w} - \iota_{\w,\z})
\bigl( a (\z-\w)_{\sip}^{\leqslant N} \, b \bigr) (\u)_{\sip} \, c
\Bigr]_{\u=\w}
\, \Bigr)_{\sip}
\,.
\]

It remains to prove that the second term resulting from the
right--hand side of~\eqref{e6.1} is equal to
\[
\Bigl(
\Bigl[ (\z^2)^{-L} \, \bigl( (\u+\z-\w)^2 \bigr)^{L} \,
(\iota_{\z,\w} - \iota_{\w,\z})
\bigl( a (\z-\w)_{\sip}^{> N} \, b \bigr) (\u)_{\sip} \, c
\Bigr]_{\u=\w}
\, \Bigr)_{\sip}
\,.
\]
This follows from the fact that
\[
\Bigl[ (\z^2)^{-L} \, \bigl( (\u+\z-\w)^2 \bigr)^{L} \,
(\iota_{\z,\w} - \iota_{\w,\z})
\bigl( a (\z-\w)_{\sip}^{> N} \, b \bigr) (\u)_+ \, c
\Bigr]_{\u=\w}
\]
contains only terms with total degree in $\z$ and $\w$ 
strictly greater than~$N$.\end{Proof}

\section{Concluding Remarks}\label{conclusions}

In this paper we develop further the theory of
vertex algebras in higher dimensions.
We start by introducing useful formal calculus techniques
including various spaces of formal series
and a formal residue functional.
This residue functional is uniquely determined
(up to a multiplicative constant) by the property that
it is translation invariant (\ref{prp2n.2}), and so it plays the role of
the integral.
In addition, it satisfies an analog of the Cauchy formula (Eq.
\eqref{eqn3nn6}).
The modes of fields can be obtained by integrating the fields
(with respect to our residue functional) against certain test functions.


Our main goal was to understand the algebraic structure obeyed
by the modes of local fields with respect to the commutator.
For this purpose we derived an analog of the Jacobi identity
for vertex algebras in higher dimensions (\ref{th-jac}).
Since the commutator of two local fields is expressed in terms of
the singular part of their operator product expansion,
a natural question arises whether
the singular parts of fields close a structure under the commutator.
Utilizing a certain degree cutoff technique
we proved that this is indeed the case (\ref{th-sp}).

Thus, if we denote by $a [\z] \, b$ the singular part $a (\z)_{\sip} b$,
we find that it closes the following structure.
The map
$a,b$ $\mapsto$ $a [\z] \, b$ $\in$
$( V \llbracket \z \rrbracket_{\z^2} )_{\sip}$ is parity preserving
and bilinear on a superspace $V$
endowed with an action of mutually commuting even endomorphisms
$T_1,\dots,T_D$,
and
the following axioms are satisfied:

\skp

(a) \hspace{2pt}(\textit{translation invariance})\hspace{2pt}
$\bigl[ \hspace{1pt} T_{\alpha},a[\z] \hspace{1pt} \bigr] \, b
= (T_{\alpha} a) [\z] \, b
= \di_{z^{\alpha}} \, a [\z] \, b$\hspace{1pt};

\skp

(b) \hspace{2pt}(\textit{skew--symmetry})\hspace{2pt}
$a [\z] \, b \, = \, (-1)^{p_a p_b} \,
\Bigl( e^{\z \spr \T} \bigl( b[-\z] \, a \bigr) \hspace{-2pt} 
\Bigr)_{\sip}$\hspace{1pt};

\skp

(c) \hspace{2pt}(\textit{Jacobi identity})\hspace{2pt}
\begin{equation*}
\begin{split}
\bigl[a& [\z] ,\, b[\w] \bigr] \, c
\\
& = \,
\Bigl(
(\z^2)^{-L} \Bigl[ \bigl( (\u+\z-\w)^2 \bigr)^{L} \,
(\iota_{\z,\w} - \iota_{\w,\z})
\bigl( a [\z-\w] \, b \bigr) [\u] \, c
\Bigr]_{\u=\w}
\, \Bigr)_{\sip}
\,
\end{split}
\end{equation*}
for $L \gg 0$,
and the expression under the substitution in the right--hand 
side belongs to the space
$(\iota_{\z,\w} - \iota_{\w,\z}) \,
V \llbracket \z,\w,\u \rrbracket_{(\z-\w)^2 \u^2}$
where setting $\u$ $=$ $\w$ makes sense.
It is expected that the obtained algebraic
structure will play in higher dimensions the same role
as vertex Lie algebras do in dimension one.

\section*{Acknowledgments}
We are grateful to E.~Frenkel, V.~G.~Kac and I.~T.~Todorov
for valuable discussions.
We thank the following institutions for their hospitality
(B.B. \& N.N.): Erwin Schr\"odinger Institute for Mathematical Physics,
(B.B.): Institut f\"ur Theoretische Physik der Universit\"at G\"ottingen,
and (N.N.): Department of Mathematics, North Carolina State University.
Bakalov was supported in part by an FRPD grant from 
North Carolina State University.
This work was done while Nikolov was visiting the
Institut f\"ur Theoretische Physik der Universit\"at G\"ottingen as
an Alexander von Humboldt research fellow.
Nikolov was partially supported
by the Research Training Network within Framework
Programme 5 of the European Commission under contract HPRN-CT-2002-00325
and by
the Bulgarian National Council for Scientific Research under contract
PH-1406.

\appendix


\section{Geometric Realization of the Residue}\label{se2.5n1}
\setcntrs

In this appendix we will provide a geometric definition of the
residue functional introduced in Sect.~\ref{se2.3n2}.
We will suppose that the space--time dimension $D$
is \textit{even}.

Let us introduce the following $1$--parameter family
$\{\MM_r\}_{r > 0}$ of $D$--dimensional real submanifolds of $\CC^D$:
\beq\label{Mr}
\MM_r :=
\bigl\{ \z \in \CC^D \; \big| \; \, \z = \lambda \, \u \,, \;\;
\lambda \in \CC \,, \;\; |\lambda| = r \,, \;\;
\u 
\in \SS^{D-1} \subset \RR^D \bigr\} \,,
\eeq
where $\SS^{D-1}$ denotes the unit sphere in $\RR^D$.
Note that $\MM_1$ is exactly the 
\textit{conformally compactified Minkowski space} (see \cite{NT,N05}).

We introduce a parameterization of $\MM_r$,
\beq\label{Mr_par}
\MM_r \, \ni \, \z = r \, e^{i\zeta} \, \u
\qquad \text{for} \quad \zeta \in \left[ 0,\pi \right)
,\ \u \in \SS^{D-1} \,,
\eeq
which shows that $\MM_r$ is diffeomorphic to
$(\SS^1 \times \SS^{D-1}) / \ZZ_2$,
with the points $(e^{i\zeta}, \u)$ and $(-e^{i\zeta}, -\u)$ 
being identified.
In particular, all $\MM_r$ are orientable for even $D$.
Thus, the volume form $dz^1 \! \wedge \dots \wedge dz^{D}$ on
$\CC^D$ can be restricted to $\MM_r$ and gives rise to a
complex measure there. 
In parameterization \eqref{Mr_par}, we have
\beq\label{vol_rest}
dz^1 \hspace{-1pt} \wedge \dots \wedge dz^{D}
\big|_{\MM_r} \, = \,
i \, r^D e^{iD\zeta} \, d\zeta \! \wedge d\sigma(\u) \,,
\eeq
where $d\sigma(\u)$ is the $\Oo (D)$--invariant volume form
$dz^1 \! \wedge \dots \wedge dz^{D} \big|_{\SS^{D-1}}$
on the unit sphe\-re~$\SS^{D-1}$.

An important property of the family $\{\MM_r\}$ is that
if $\z \in \MM_r$ and $\w \in \MM_{r'}$ for $r \neq r'$
then $(\z - \w)^2 \neq 0$. Indeed, writing
\beq\label{mr-p1}
\z = r \, e^{i\zeta} \, \u \,, \;\; 
\w = r' \, e^{i\zeta'} \, \u'\,, \quad
\u \spr \u' = \cos\al = (e^{i\al} + e^{-i\al}) / 2 
\,,
\eeq
we find that
\beq\label{mr-p2}
(\z - \w)^2 \, = \,
\bigl( r \, e^{i\zeta} - r' \, e^{i(\zeta'+\al)} \bigr)
\bigl( r \, e^{i\zeta} - r' \, e^{i(\zeta'-\al)} \bigr) \,.
\eeq
This shows that for $n \in \ZZ$ the formal Taylor expansion 
$\iota_{\z,\w} \hspace{1pt} ((\z - \w)^2)^n$ 
(see \eqref{eq2.21n2}) in the above parameterization
corresponds to a geometric series expansion for $r>r'$.
Note also that the conformal inversion
$\z \mapsto \z / \z^2$ maps $\MM_r$ onto $\MM_{r^{-1}}$.

\skp

\begin{Proposition}\label{prp_geom-res}
For\/ $f (\z) \in \CC \left[ \z \right]_{\z^2}$ and\/
$g (\z,\w) \in \CC \left[ \z,\w \right]_{\z^2\w^2(\z - \w)^{2}}$,
we have{\rm:}
\begin{align}
\label{gi1}
& \Res_{\z} \, f (\z) \, = \,
(i \hspace{1pt} \pi \hspace{1pt} \mathcal{V}_{D-1})^{-1} \,
\mathop{\int}\limits_{\hspace {-4pt}\MM_r} \,
f (\z) \ dz^1 \hspace{-1pt} \wedge \dots \wedge dz^{D}
\,,
\\ \label{gi2}
& \Res_{\z} \ \iota_{\z,\w} \, g (\z,\w)
\\ \notag
& \qquad = \,
(i \hspace{1pt} \pi \hspace{1pt} \mathcal{V}_{D-1})^{-1} \,
\mathop{\int}\limits_{\hspace {-4pt}\MM_r} \,
g (\z,\w) \ dz^1 \hspace{-1pt} \wedge \dots \wedge dz^{D}
\quad\text{for}\quad 
\w \in \MM_{r'},\ r' < r \,,
\\ \label{gi3}
& \Res_{\z} \ \iota_{\w,\z} \ g (\z,\w)
\\ \notag
& \qquad = \,
(i \hspace{1pt} \pi \hspace{1pt} \mathcal{V}_{D-1})^{-1} \,
\mathop{\int}\limits_{\hspace {-4pt}\MM_r} \,
g (\z,\w) \ dz^1 \hspace{-1pt} \wedge \dots \wedge dz^{D}
\quad\text{for}\quad 
\w \in \MM_{r'},\ r' > r \,,
\end{align}
where\/ $\mathcal{V}_{D-1}
= \mathop{\displaystyle \int}\limits_{\hspace {-4pt}\SS^{D-1}} d\sigma(\u)$.
\end{Proposition}

\skp

\begin{Proof}
It is enough to check (\ref{gi1}) for the functions
$(\z^2)^n \, h(\z)$, where $n \in \ZZ$ and $h(\z)$ is a
harmonic homogeneous polynomial of degree $m$.
Then \eqref{gi1} follows from \eqref{vol_rest} and the formulas
\[ 
\mathop{\int}\limits_{\SS^{D-1}} \,
h(\u) \, d\sigma(\u) \, = \, \de_{m,0} \, h(0) \, \mathcal{V}_{D-1} \,,
\qquad
\int\limits_0^\pi \,
e^{i D \ze +  2 i n \ze} \, d\ze \, = \, \pi \, 
\de_{n,-\frac{D}2} \;.
\]
Equation (\ref{gi2}) follows from~(\ref{gi1}) because the expansion
$\iota_{\z,\w} \, g (\z,\w)$
converges uniformly to $g (\z,\w)$ for $\z \in \MM_r$,
$\w \in \MM_{r'}$ and fixed $r > r'$ 
(see \eqref{mr-p2}).
Finally, (\ref{gi3}) is proved in the same way as (\ref{gi2})
but for $r < r'$.\end{Proof}

\skp

\bibliographystyle{amsalpha}


\end{document}